\documentclass{aa}  

\usepackage{graphicx}
\usepackage{pdflscape}
\usepackage{txfonts}
\usepackage{natbib}
\usepackage{color}
\usepackage{amssymb}
\usepackage{mathrsfs}
\usepackage{multirow}

\begin{document} 

   \title{Gas dynamics in the inner few AU around the Herbig B[e] star \\MWC297 \thanks{Based on observations made with ESO Telescopes at the La Silla Paranal Observatory under programme IDs 081.D-0230, 083.C-0590, 089.C-0959, and 089.C-0563.}}

   \subtitle{Indications of a disk wind from kinematic modeling and velocity-resolved\\ interferometric imaging}
   
   \titlerunning{Gas dynamics in the inner few AU around the Herbig B[e] star MWC297}

   \author{
    Edward Hone\inst{1}
     \and
    Stefan Kraus\inst{1}
     \and
    Alexander Kreplin\inst{1}
     \and
    Karl-Heinz Hofmann\inst{2}
     \and
    Gerd Weigelt\inst{2}
     \and
    Tim Harries\inst{1}
     \and
    Jacques Kluska\inst{1}
    }
    
   \authorrunning{Hone et al.}
    
   \institute{Department of Physics and Astronomy, 
            University of Exeter, Stocker Road, Exeter, EX4 4QL, UK
            \and
            Max-Planck Institut f\"ur Radioastronomie, Auf dem H\"ugel 69, 53121 Bonn, Germany
              }

   \date{}

 
  \abstract
    {}
    {Circumstellar accretion disks and outflows play an important role in star formation.
    By studying the continuum and Br$\gamma$-emitting region of the Herbig~B[e] star MWC297 with high-spectral- and high-spatial resolution we aim to gain insight into the wind-launching mechanisms in young stars.}
    {We present near-infrared AMBER ($R=12,000$) and CRIRES ($R=100,000$) observations of the Herbig B[e] star MWC297 in the hydrogen Br$\gamma$-line.
    Using the VLTI unit telescopes, we obtained a $uv$-coverage suitable for aperture synthesis imaging.
    We interpret our velocity-resolved images as well as the derived two-dimensional photocenter displacement vectors, and fit kinematic models to our visibility and phase data in order to constrain the gas velocity field on sub-AU scales.
    }
    { 
    The measured continuum visibilities constrain the orientation of the near-infrared-emitting dust disk, where we determine that the disk major axis is oriented along a position angle of ${\sim99.6} \pm 4.8^\circ$.
    The near-infrared continuum emission is $\sim 3.6\times$ more compact than the expected dust-sublimation radius, possibly indicating the presence of highly refractory dust grains or optically thick gas emission in the inner disk.
    Our velocity-resolved channel maps and moment maps reveal the motion of the Br$\gamma$-emitting gas in six velocity channels, marking the first time that kinematic effects in the sub-AU inner regions of a protoplanetary disk could be directly imaged.
    We find a rotation-dominated velocity field, where the blue- and red-shifted emissions are displaced along a position angle of $24^\circ\pm3^\circ$ and the approaching part of the disk is offset west of the star.
    The visibility drop in the line as well as the strong non-zero phase signals can be modeled reasonably well assuming a Keplerian velocity field, although this model is not able to explain the $3\sigma$ difference that we measure between the position angle of the line photocenters and the position angle of the dust disk.
    We find that the fit can be improved by adding an outflowing component to the velocity field, as inspired by a magneto-centrifugal disk-wind scenario.
    }
    {This study combines spectroscopy, spectroastrometry, and high-spectral dispersion interferometric, providing yet the tightest constraints on the distribution and kinematics of Br$\gamma$-emitting gas in the inner few AU around a young star.
    All observables can be modeled assuming a disk wind scenario.
    Our simulations show that adding a poloidal velocity component causes the perceived system axis to shift, offering a powerful new diagnostic for detecting non-Keplerian velocity components in other systems.
    }

\keywords{stars: formation --
stars: circumstellar matter --
stars: variables: Herbig Ae/Be
ISM: jets and outflows --
ISM: individual objects: MWC297 --
techniques: interferometric --
techniques: high angular resolution
}

   \maketitle
%

\section{Introduction}
\label{Sec:intro}

The formation of stars is characterized by the presence of circumstellar disks and outflows.
These outflows play a vital role in the star formation process, removing excess angular momentum from the inner circumstellar environment, impacting the surrounding interstellar medium, and influencing the next generation of star formation.
We observe these phenomena around young, low-mass T\,Tauri stars and their intermediate-mass counterparts, the Herbig Ae/Be stars.\\

Various models have been proposed to explain how outflows might be launched from these young stellar objects.
Outflows could be produced by magneto-centrifugally driven winds from the accretion disk \citep{1982Blandford}, as stellar winds accelerated along open magnetic field lines anchored to the stellar surface \citep{2005Matt} or from a region where the stellar magnetic field interacts with the accretion disk \citep{1994Shu}.
The work of \citet{2006Ferreira} outlines how these three mechanisms can combine in the inner regions of protoplanetary disks but their individual contributions are still under debate \citep[see models in][2016]{2014Tambovtseva}.\\
Distinguishing between these models observationally is extremely challenging, requiring us to spatially resolve the jet-launching region at scales of just a few astronomical units (au) as well as achieving the high spectral resolution required to resolve the kinematics in outflow-tracing spectral lines.
Spectrally dispersed interferometry allows us to achieve the required resolution to spatially and spectrally resolve the kinematics traced by the Br$\gamma$ line at 2.166\,$\mu$m in the K-band, which is emitted by the hot gas in the wind or jet-launching region.
The Br$\gamma$ emission appears to trace different processes in different stars, such as Keplerian rotation \citep{2012Kraus1}, disk winds \citep{2011Weigelt,2015GarciaLopez,2015CarrattioGaratti,2016Kurosawa} and magnetospheric accretion \citep{2010Eisner}.
The work of \citet{2008Kraus} showed that, for a selection of five Herbig Ae/Be stars, the radial extension of the Br$\gamma$ line-emitting region relative to the continuum differs dramatically, suggesting that in different objects the physics traced in the line are significantly varied.\\

\object{MWC297} is one of the nearest Herbig stars, with a mass of ${\sim}10M_{\odot}$, spectral type B1.5V and a distance of $170$\,pc \citep{2015Fairlamb}.
Spectroscopic observations by \citet{1997Drew} indicate that MWC297 is rapidly rotating with ${v\sin\,i\,=\,350\,\pm\,50\text{\,kms}^{-1}}$.
Stellar rotation becomes critical when ${v\,=\,v_{crit}\,=\,\sqrt{2GM_{*}/(3R_{*})}}$ \citep{2000Maeder} which for MWC297 is ${450\,\pm\,50\text{\,kms}^{-1}}$ (assuming ${R_{*}\,=\,6\,R_{\odot}}$ and ${M_{*}\,=\,10\,M_{\odot}}$, \citealt{1997Drew,2011Weigelt}).
Assuming that the disk and star are aligned, this rapid rotation is called into question by the interferometric observations of the disk inclination, which was determined to be ${\sim}20^\circ$ by \citet{2007Malbet}, \citet{2008Acke} and \citet{2011Weigelt}. 
With these parameters, $v$ takes the value ${{\sim}1023\,\text{kms}^{-1}}$ which far exceeds the ${450\text{\,kms}^{-1}}$ break up velocity. This `unphysical' result could point to a possible misalignment between the star and the disk.\\

Important insights into the au-scale geometry of the disk around MWC297 have been obtained with near- and mid-infrared interferometry \citep{2007Malbet, 2008Acke, 2008Kraus, 2011Weigelt}.
An interesting recurring result from these works is that the observed continuum radius is much smaller than the predicted dust sublimation radius of ${\sim}3\,$au (assuming a dust sublimation temperature for silicate grains of $T_{\mathrm{sub}} = 2000$\,K).
This characteristic has also been found in other luminous objects, such as \object{Z\,CMa} and \object{V1685 Cyg} \citep{2005Monnier}.
Like MWC297 (B1), both of these objects are early-type B stars (Z\,CMa: B0, \citealt{2004vandenAncker}; V1685 Cyg: B3, \citealt{2004Hernandez}), indicating that this could be characteristic for young, luminous Herbig Be stars.\\

MWC297 was also resolved with VLTI/PIONIER in the $H$-band by \citet{2017Lazareff}.
They fitted Gaussian-modulated ring models to their visibility and closure phase data but could not find any evidence for a sharp inner edge around MWC297.
Instead, the measured visibility profile is very smooth and indicated a Gaussian-like profile with the system axis along a position angle (PA) of $13.7\pm 1.7^\circ$, which is consistent with our PA estimate within $0.88\sigma$.
The morphology of the near-infrared continuum varies for different objects, with lower-mass Herbig~Ae stars showing a ring shape (tracing the inner dust rim) and the higher-mass Herbig~Be stars having a more radially extended ring structure, sometimes leading to Gaussian-shaped continuum emission \citep{2017Lazareff}.
\object{MWC297} occupies the more massive, luminous end of the Herbig Be star spectrum, with a spectral type of B1.5V and an estimated mass of $10M_\odot$ , but it displays a very intriguing continuum geometry with the NIR emission well inside the dust sublimation radius and no "hole" in the emission indicative of a ring-like structure.
The variation of optical depth in the inner gas disk was investigated by \citet{2004Muzerolle}, who found that for high accretion rates ($\dot{M} \gtrsim 10^{-8}$ M$_{\odot}$yr$^{-1}$) the inflowing gas can become optically thick.
Further study of the accretion rate and gas opacity in the inner disk is beyond the scope of this study.\\

Spectrally dispersed interferometry with VLTI/AMBER offers a unique opportunity to observe wavelength-dependent changes in the structure at medium spectral resolution ($R=1,500$) or the kinematics of line emitting gas at high spectral resolution ($R=12,000$), retaining the extremely high angular resolution achieved with optical interferometry.
This technique was employed by both \citet{2007Malbet} and \citet{2011Weigelt} to observe the Br$\gamma$ emission line of MWC297.
\citet{2007Malbet} simulated their single baseline $R=1,500$ data with a model comprised of two codes: one simulating an optically thick, continuum-emitting disk around MWC297, and the other modeling a stellar wind from the central star simulating the source of the line emission.
Their model provides a good fit to the observed spectral energy distribution (SED), differential visibility and Br$\gamma$ spectrum, but with no differential phase measurements and a very small sample of data (just one baseline), the validity of their kinematic model could not be determined.
In addition, they assumed a double-peaked Br$\gamma$ line profile for their kinematic modeling, while a later reanalysis of the same ISAAC data set showed that the double-peaked profile was a calibration artifact and that the line profile is single-peaked.\\

\citet{2011Weigelt} developed a more complex kinematic model which included a continuum-emitting ring with an extended disk wind component contributing to the Br$\gamma$ line emission.
The geometry of the disk wind component of their model is similar to that detailed in the work of \citet{2006Kurosawa} with the parameterization of mass-loss and acceleration detailed further in \citet{2011Weigelt}.
The shape of the differential visibilities and phases from the model are a good match to the observed ones, but there are significant residuals between the model and the data which leaves room for further study into the object with similar techniques.
The data set that they modeled contained only a small range of short baselines (14 to 42m) along a linear configuration with a position angle of $68^\circ$, which is insufficient to constrain the two-dimensional (2D) geometry of the line-emitting region.\\

In this paper we use a richer spectrally dispersed interferometry data set to spatially and spectrally resolve the Br$\gamma$ line of MWC297 and study the distribution and kinematics of the line-emitting gas in order to place physical constraints on the processes traced by Br$\gamma$.
In Sect. \ref{Sec:obs} we present AMBER spectro-interferometry and CRIRES spectro-astrometric observations which we interpret in Sect. \ref{Sec:photocenter} by measuring the 2D photocenter displacements across the Br$\gamma$ line and in Sect. \ref{Sec:2Dmodel} by fitting the visibilities with geometric models.
In Sect. \ref{Sec:kinematicmodel} we present kinematic models of both Keplerian rotation and a magneto-centrifugally driven disk wind and compare the models to our observed data.



\section{Observations and data reduction}
\label{Sec:obs}

\subsection{VLTI/AMBER spectro-interferometry}

\begin{figure}
        \includegraphics[width=\columnwidth]{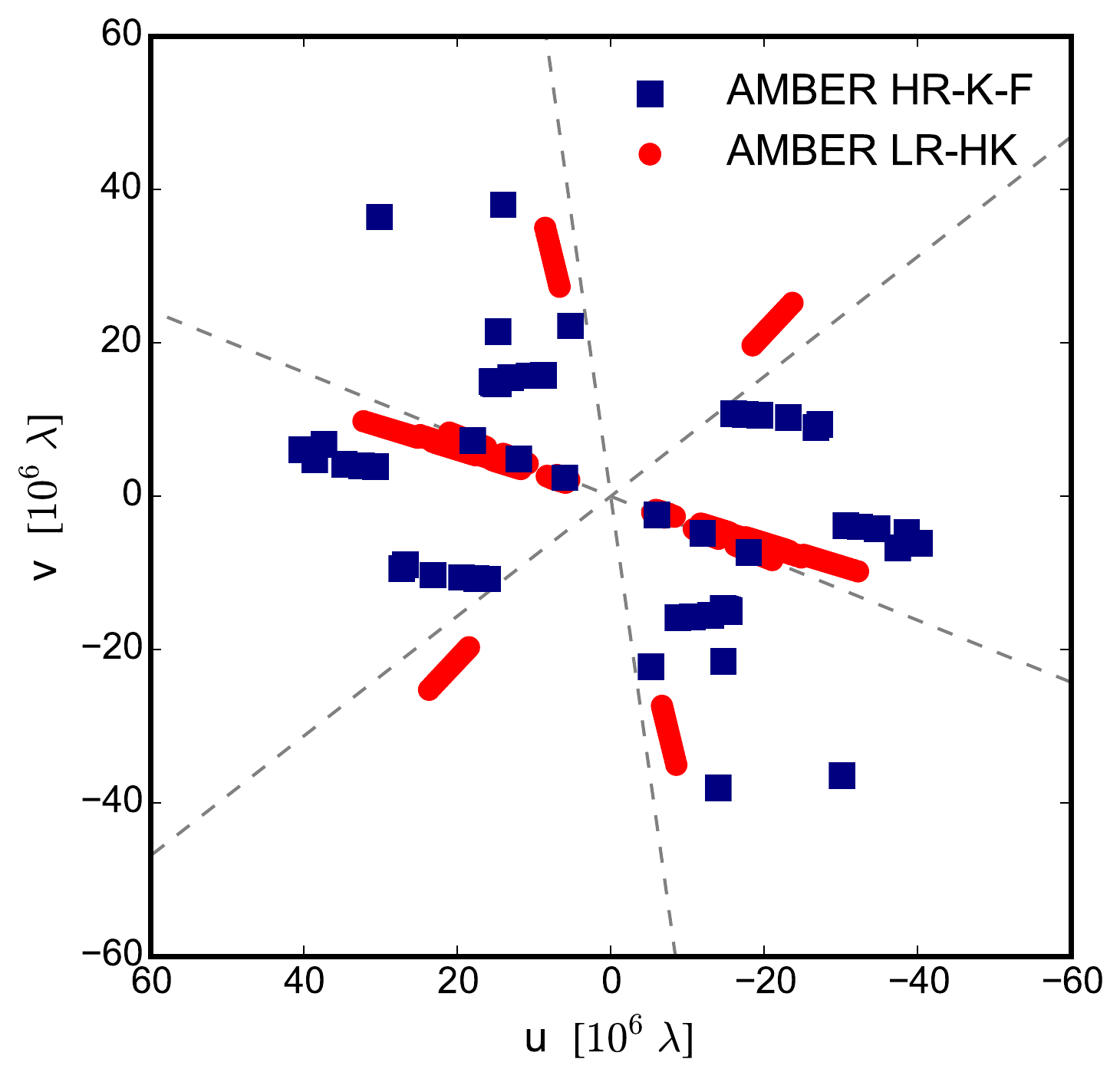}
    \caption{$uv$-coverage achieved with our VLTI/AMBER interferometric observations of MWC297 for both $R=30$ (red circles) and $R=12,000$ (blue squares) spectral resolutions. The gray dashed lines indicate the slit orientations for our CRIRES spectro-astrometric observations.}
    \label{fig:uvcov}
\end{figure}

\begin{table*}
\small
\caption{Observation log for our MWC297 data taken with VLTI/AMBER and VLT/CRIRES.}
\label{obslog}
\centering
\vspace{0.1cm}
\begin{tabular}{c c c c c c c c c c}
\hline\hline
Instrument & Date & Telescopes & UT  & DIT & NDIT & Proj. baselines & PA & Calibrator & UD diam.\tablefootmark{a}\\ 
& & & [h:m] & [ms] & \# &  [m] & [$\circ$] & &[mas] \\  \hline 
AMBER & 2008-04-06 &H0/G0/E0 & 08:14 & 8000 & 100 & 14.0\,/\,28.0\,/\,42.0 & 247.9\,/\,247.9\,/\,247.9 & \object{HD175583} & $1.00 \pm 0.10$ \\ 
HR-K & 2009-05-13 &U2/U3/U4 & 10:01 & 200 & 1200 & 46.0\,/\,48.0\,/\,75.8  &45.9\,/\,118.5\,/\,83.1 & \object{HD187660} & $1.85 \pm 0.19$ \\ 
" & 2009-06-13 &U2/U3/U4 & 04:15 & 1000 & 560 & 41.0\,/\,61.1\,/\,82.2  &34.0\,/\,108.5\,/\,79.7 & \object{HD166460} & $1.54 \pm 0.15$ \\ 
" & " &U2/U3/U4 & 08:22 & 3000 & 125 & 44.5\,/\,45.4\,/\,71.0  &121.3\,/\,45.5\,/\, 83.0 & \object{HD187660} & $1.85 \pm 0.19$ \\ 
" & " &U2/U3/U4 & 08:38 & 500 & 600 & 41.9\,/\,44.9\,/\,67.1  &123.8\,/\,45.1\,/\,82.9 & \object{HD187660} & $1.85 \pm 0.19$ \\ 
" & 2009-08-03 &U2/U3/U4 & 03:46 & 3000 & 200 & 46.6\,/\,54.9\,/\,84.3  &45.8\,/\,113.9\,/\,83.0 & \object{HD166460} & $1.54 \pm 0.15$ \\ 
" & 2012-04-03 &U1/U2/U3 & 08:19 & 500 & 840 & 39.1\,/\,49.4\,/\,87.7  &28.8\,/\,13.1\,/\,20.0 & \object{HD152040} & $0.57 \pm 0.06$ \\ 
" & 2012-05-04&U1/U2/U3 & 10:02 & 400 & 1200 & 46.5\,/\,56.4\,/\,102.4  &46.0\,/\,34.3\,/\,39.6 & \object{HD166460} & $1.54 \pm 0.15$ \\ 
" & 2012-06-01 &U2/U3/U4 & 05:57 & 500 & 960 & 43.9\,/\,62.4\,/\,88.3  &220.0\,/\,109.0\,/\,81.4 &\object{HD152040} & $0.57 \pm 0.06$\\ 
\hline
AMBER & 2007-04-14 & H0/G0/E0 & 08:09 & 50 & 5000 & 28.2\,/\,42.2\,/\,14.1 & 68.1\,/\,68.1\,/\,68.1 & \object{HD166460} & $1.54 \pm 0.15$ \\
LR-K & 2007-06-19 & H0/G0/E0 & 07:10 & 50 & 5000 & 30.1\,/\,45.2\,/\,15.1 & 73.1\,/\,73.1\,/\,73.1 & \object{HD166460} & $1.54 \pm 0.15$ \\
" & 2007-09-01 & H0/G0/E0 & 00:46 & 100 & 10000 & 31.9\,/\,47.9\,/\,16.0 & 72.5\,/\,72.5\,/\,72.5 & \object{HD172051} & $0.64 \pm 0.06$ \\
" & 2007-09-02 & H0/G0/E0 & 00:23 & 100 & 2500 & 31.7\,/\,47.6\,/\,15.9 & 72.1\,/\,72.1\,/\,72.1 & \object{HD160213} & $0.15 \pm 0.01$ \\
" & " & H0/G0/E0 & 00:36 & 50 & 5000 & 31.9\,/\,47.9\,/\,16.0 & 72.4\,/\,72.4\,/\,72.4 & \object{HD160213} & $0.15 \pm 0.01$ \\
" & 2008-06-03 & H0/G0/E0 & 08:18 & 50 & 15000 & 29.7\,/\,44.6\,/\,14.9 & 73.0\,/\,73.0\,/\,73.0 & \object{HD177756} & $0.51 \pm 0.05$ \\
" & 2009-05-16 & G1/D0/H0 & 08:21 & 100 & 5000 & 64.8\,/\,68.1\,/\,63.4 & 137.3\,/\,14.2\,/\,73.1 & \object{HD166460} & $1.54 \pm 0.15$ \\
\hline
CRIRES & 2009-09-12 & UT1 & 01:43 & 1500 & 1500 & N/A & 8.0\,/\,68.0\,/\,128.0 & \object{HD165185} & $0.56 \pm 0.06$ \\
\hline \hline
\end{tabular}
\tablefoot{$^{a}$The calibrator UD diameter (\textit{K} band) was taken from the JMMC Stellar Diameters Catalogue \citep{2010Lafrasse}.}
\end{table*}

We observed MWC297 between 2008 and 2012 with ESO's Very Large Telescope Interferometer (VLTI) and its beam combination instrument AMBER \citep{2007Petrov}.
Our VLTI/AMBER observations are outlined in Table \ref{obslog} and the $uv$-coverage of these observations is shown in Fig. \ref{fig:uvcov}.
The data set that we present contains the three baselines from the linear configuration of \citet{2011Weigelt} obtained with the 1.8\,m Auxiliary Telescopes (ATs) along a PA of $67^\circ$ but we improve on this $uv$-coverage by adding nine additional observations with the 8.2\,m Unit Telescopes (UTs).
Our additional AMBER data contain baselines with position angles ranging from ${\sim}10^\circ$ to ${\sim}130^\circ$ and the full possible range of baseline lengths achievable with the UTs (30-132\,m).
In order to improve the signal-to-noise ratio (SNR), we employed the fringe-tracking instrument FINITO \citep{2004Gai, 2008LeBouquin}.
The use of long integration times (up to 8000\,s) improves the SNR on the wavelength-differential observables, but can also bias the absolute visibility calibration.
Therefore, we calibrate the continuum visibility level using low-spectral-resolution AMBER data.
Each science observation of MWC297 was accompanied by a calibrator observation, with the targets \object{HD175583}, \object{HD187660}, \object{HD166460}, \object{HD152040} and \object{HD164259} used as interferometric calibrators (detailed in Table \ref{obslog}).
The data was reduced with our own AMBER data-processing software package, which uses the pixel-to-visibility matrix algorithm P2VM \citep{2007Tatulli, 2009Chelli} in order to extract visibilities, differential phases, and closure phases for each spectral channel of an AMBER interferogram.
We adopt heliocentric line-of-sight velocities for each observation using the ESO Airmass tool.

\subsection{VLT/CRIRES spectro-astrometry}

We complement our AMBER spectro-interferometric data with observations taken with the CRIRES spectrograph at the VLT \citep{2004Kaeufl}.
The observation details are outlined in Table \ref{obslog}.
Spectro-astrometry uses long slit spectra to measure the wavelength-dependent centroid offset of an unresolved object with respect to the continuum.
Whilst this does not let us formally resolve the object spatially, it offers extremely high spectral resolution of $R=100,000$, meaning that we can fully resolve the detailed structure in the Br$\gamma$-line profile.
The centroid position can be measured with much higher precision than the size of the point-spread-function (PSF), allowing us to measure the small-scale photocenter displacements characteristic of the kinematics in the inner disk.
By measuring the centroid offset along three slits oriented at position angle offsets of 60$^\circ$ we can derive 2D photocenter displacement vectors.
We chose to observe with slits oriented at PAs of $8^\circ$, $68^\circ$, and $128^\circ$ (and their corresponding anti-parallel PAs $188^\circ$, $248^\circ$ and $308^\circ$), in order to complement the interferometric observations published by \citet{2011Weigelt} that were obtained along a baseline PA of $68^{\circ}$.
Our CRIRES observations were obtained on 2009 September 12 using an integration time of 1.5s and a slit width of 0.197".\\
To derive the differential phase signal from our CRIRES observations in order to later model our AMBER and CRIRES data simultaneously, we follow the method outlined by \citet{2012Kraus3}.
This method makes use of the fact that the centroid shifts $X(v)$ that we calculate above are mathematically equivalent to the differential phase signals that we measure using spectro-interferometry at very short baseline lengths.
We employ the equation $\phi = -\frac{2\pi X(v)}{\sigma}$ to determine the differential phase $\phi$ from the observed centroid shifts $X(v)$ for each spectral channel $v$, where $\sigma$ is the full-width at half-maximum (FWHM) of the PSF measured in the spectrum.


\section{Geometric modeling}
\label{Sec:2Dmodel}


\begin{figure*}
\centering
\includegraphics[width=18.4cm]{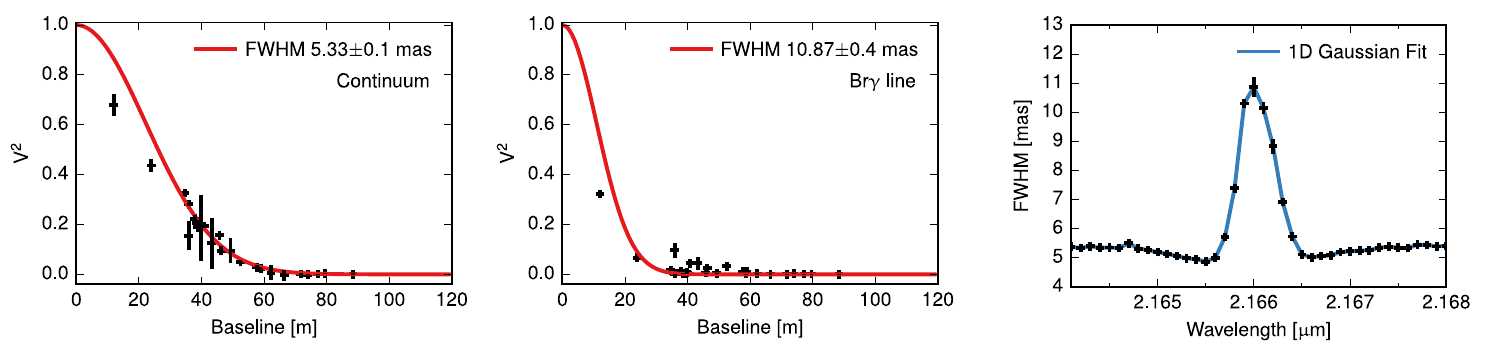}
\caption{
Left: Results of our 1D Gaussian model fit to deprojected visibilities for continuum channels (Error rounded up to 0.1mas).
Center: Results of our 1D Gaussian model fit to deprojected visibilities for the central channel in the Br$\gamma$ line.
Right: Wavelength-dependent 1D Gaussian fit FWHM diameters for each channel across the Br$\gamma$ line.
}
\label{fig:linevis}
\end{figure*}

\subsection{AMBER LR-K continuum visibilities}
\label{Sec:2DmodelCont}

To measure the characteristic size and morphology of the near-infrared continuum disk around MWC297 we fit a 2D geometric model to our low-spectral-resolution (R=30) AMBER K-band observations.
These observations are outlined in Table \ref{obslog} and provide a good coverage towards position angles $14^\circ$, $137^\circ$, and a range between $68^\circ$ and $73^\circ$ (Fig. \ref{fig:uvcov}, red data points).
We fit a 2D geometric model of an elongated Gaussian brightness distribution to the data.
Our best fit model parameters indicate a FWHM size of $4.9 \pm 0.1$\,mas with the minor axis (later referred to as the `system axis') along position angle of $9.6^\circ \pm 4.8^\circ$.
We find that the FWHM sizes of the minor and major axes are $4.9 \pm 0.1$ mas and $4.2 \pm 0.1$ mas respectively, corresponding to an inclination of $32^\circ \pm 3^\circ$.\\
Our derived continuum size is in good agreement with the results derived by \citet{2008Acke} and \citet{2011Weigelt}, who measured characteristic continuum FWHM sizes of ${\sim}4.3$\,mas and ${\sim}4.6$\,mas, respectively.
Additionally, our inclination measurements are in agreement with the tentative disk inclination of ${\sim}20^\circ$ with an upper limit of $40^\circ$ determined by \citet{2008Acke}.
Below we  use the orientation and basic geometry of the continuum disk to compare with the distribution of the line-emitting gas.\\


\subsection{AMBER HR-K line visibilities}

Observing with high-spectral-resolution interferometry allows us to obtain visibility information for individual spectral channels across the Br$\gamma$ line, thus constraining the spatial distribution of the line-emitting gas.
The differential visibility signatures show a strong decrease in the line, indicating that the line-emitting region is more extended than the continuum, and so we apply our geometric modeling code to fit the de-projected visibilities channel-by-channel across the Br$\gamma$ line, applying the equation $r_{uv\theta i} = \sqrt{u_{\theta}^2 + v_{\theta}^2\cos(i)^2}$ \citep{2007Berger} where $u$ and $v$ are the spatial co-ordinates of the baseline vector, $\theta$ is the disk PA, and $i$ is the disk inclination.\\

By applying this transformation for all spectral channels we take the approximation that the line emission is seen under the same inclination as the continuum emission.
We apply this de-projection to our baseline vectors, adopting the position angle and inclination values of $9.6^\circ$ and $32^\circ$ determined from modeling our AMBER LR data (Sect.~\ref{Sec:2DmodelCont}).
The differential visibility data is then split by wavelength in order to derive the baseline-dependent visibilities for each spectral channel.
We fit a simple 1D Gaussian model to the deprojected baseline-dependent visibility data across the Br$\gamma$ line and determine the wavelength-dependent FWHM of the line-emitting region (Fig.~\ref{fig:linevis}).
There is a clear size increase detected up to a peak FWHM size of $10.87{\pm0.4}$\,mas in the line center, similar to the wavelength-dependent size profile shown in \citet{2011Weigelt}, where the line-emitting region increased in size towards the line center to a peak FWHM size of 12.61\,mas.


\section{Photocenter analysis}
\label{Sec:photocenter}


\subsection{CRIRES spectro-astrometry}
\label{subsec:criresp}

\begin{figure*}
\centering
\includegraphics[width=14cm]{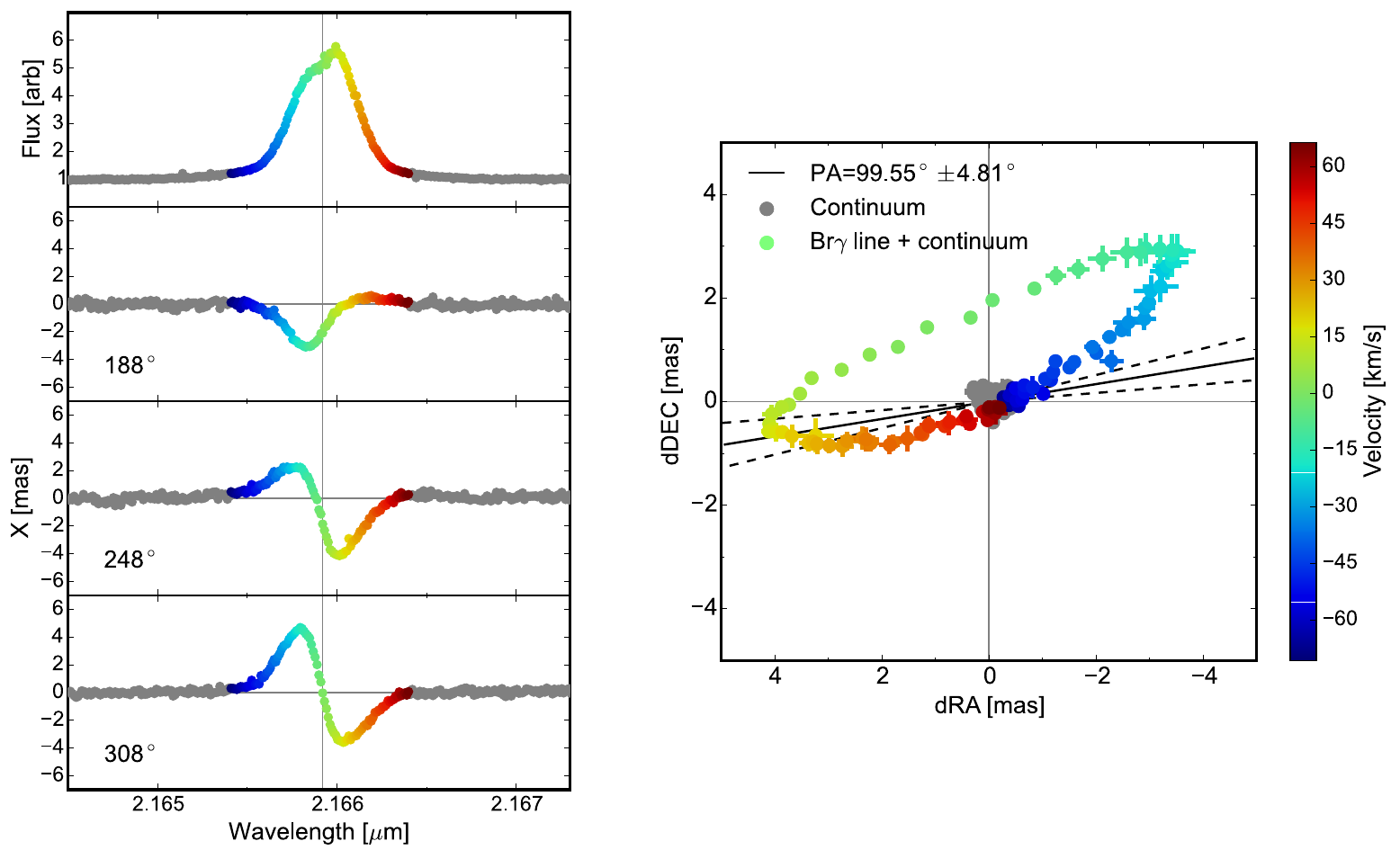}
\caption{Left: VLT/CRIRES high-resolution ($R=100,000$) Br$\gamma$ spectrum (top panel) and centroid displacements of the Br$\gamma$-emitting region (lower panels) for MWC297. The position angle of each slit is shown to the left of each phase panel.
Right: Derived 2D photocenter displacement vectors of the Br$\gamma$ + continuum emission for MWC297. The black line shows the position angle of the disk major axis determined from our AMBER LR observations.}
\label{fig:MWC297-CRIRES}
\end{figure*}

CRIRES spectro-astrometry allows us to directly measure centroid displacements between spectral channels at very high spectral resolution with high SNR, making it a powerful tool for determining the nature of the kinematics traced in spectral lines.
We calculate the photocenter displacement vectors ($\vec{p}$) for the differential phases calculated from our CRIRES spectro-astrometric data by combining the spectro-astrometric signal for each of our three observed slit vectors (corresponding to the baseline vectors $B_i$) and finding the best-fit photocenter vector using the Nelder-Mead simplex algorithm \citep[as described in][]{2012Kraus3}.
Using this method we obtain the series of wavelength-dependent photocenter displacement vectors of the combined continuum and line emission shown in Figure \ref{fig:MWC297-CRIRES} (right panel).\\

The derived photocenter vectors show an offset between the red-shifted and blue-shifted vectors along a position angle of $114\pm3^\circ$.
The photocenters of the higher-velocity channels are less displaced from the continuum emission than the lower-velocity channels, indicative of a rotational field where the velocity decreases with radius, such as Keplerian rotation.
We fit a linear function to these photocenter displacement vectors in order to determine the axis along which the blue and red-shifted lines are displaced. We determine that this "axis of motion" is oriented along PA $114^\circ\pm3^\circ$, which differs from the PA of the disk major axis of $99.6^\circ \pm 4.8^\circ$ determined from our AMBER-LR observations.
The photocenter shifts for the whole line exhibit an interesting arc pattern, with the vectors becoming more displaced to North as they approach the line center.
It is possible that this pattern is caused by obscuration of the further regions of the disk shifting the photocenter vectors towards the observer, an idea that we explore further in Sect.~\ref{Sec:kinematicmodel} as part of our modeling of the disk kinematics.\\

Various papers disagree on the shape of the Br$\gamma$ line profile of MWC297.
The VLT/ISAAC spectrum shown by \citet{2006GarciaLopez} and \citet{2007Malbet} indicates a double-peaked structure but this is in contrast with later work, in particular the re-reduction of the ISAAC data set presented in \citet{2008Kraus} and the AMBER HR-K data set by \citet{2011Weigelt}.
Our very-high-spectral-resolution data taken with the VLT/CRIRES instrument (Figure \ref{fig:MWC297-CRIRES}) show that the Br$\gamma$-line from MWC297 is single-peaked, although these data are from a later epoch than the 2004 ISAAC data.
Some small asymmetries seen in the CRIRES Br$\gamma$ line profile are consistent with the line profiles shown for a disk wind at low inclinations in the work of \citet{2016Tambovtseva}.

\subsection{AMBER spectro-interferometry}
\label{subsec:amberp}

\begin{figure*}[htbp]
\centering
\includegraphics[width=18.4cm]{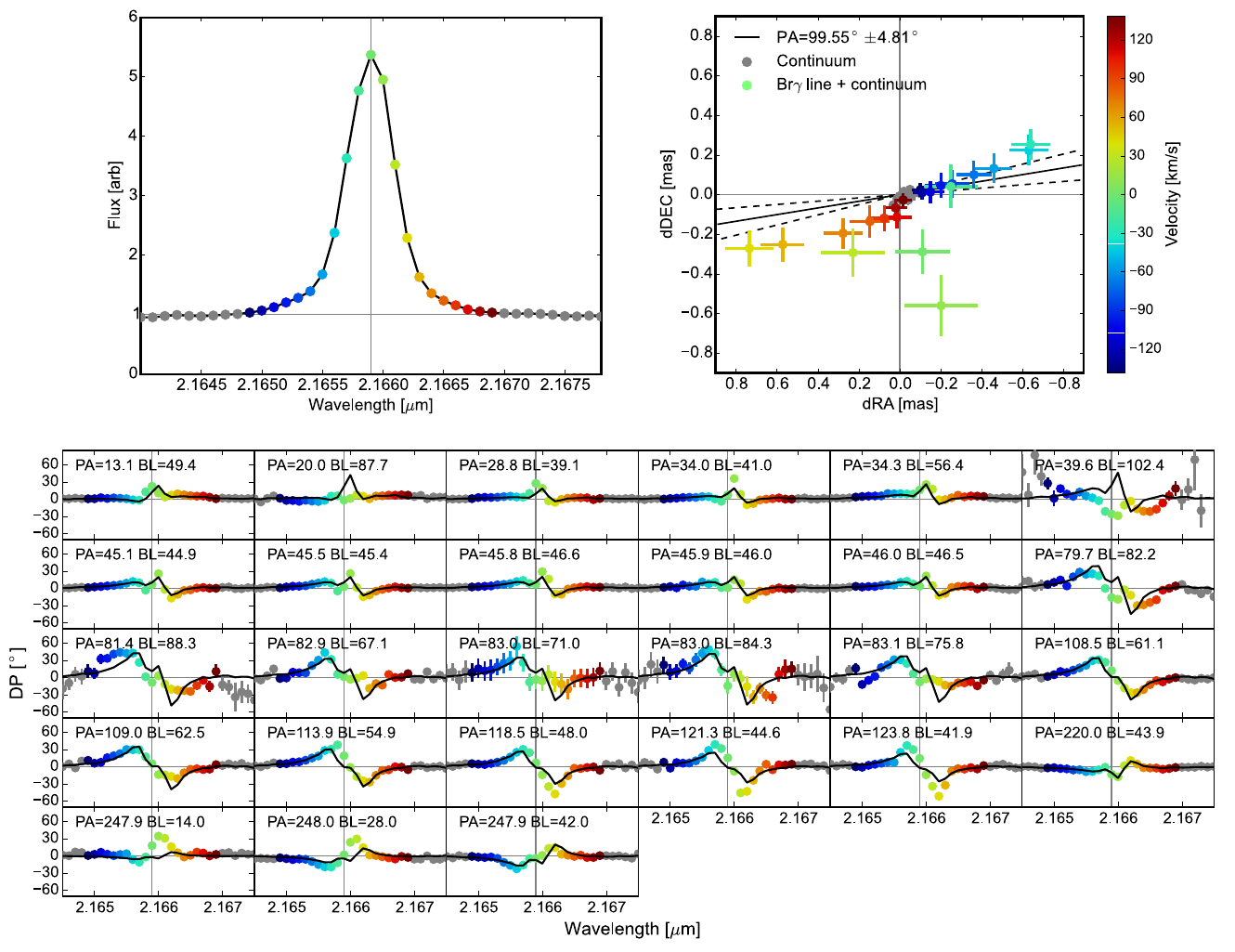}
\caption{
Upper left panel: K-band spectrum of MWC297 measured with AMBER, averaged across all baselines.
Upper right panel: Derived 2D photocenter displacement vectors for the Br$\gamma$ line + continuum determined from our AMBER HR-K data. The black line shows the position angle of the disk major axis determined from our AMBER LR observations.
Lower panel: Differential phases for MWC297 observed with AMBER (circular points) compared with differential phases corresponding to the photocenter shifts shown in the upper right panel of this Figure (black lines). The frame with PA=39.6, BL=102.4 is offset by $180^{\circ}$ in order to see the differential phase changes at the Br$\gamma$ line center.
}
\label{fig:photocenter}
\end{figure*}

Measuring differential phases allows us to gain unique insight into small-scale spatial displacements between spectral channels.
These displacements are a very powerful diagnostic for the gas kinematics at scales of just a few stellar radii.
Initially we calculate the photocenter displacement vectors $\vec{p}$ (of both the continuum and line-emitting region with respect to the continuum center) from our AMBER differential phase measurements by solving the set of 2D equations:

\begin{equation}
\vec{p} = -\frac{\phi_i}{2\pi}\cdot \frac{\lambda}{\vec{B_i}},
\label{eq:photocenter}
\end{equation}

\noindent where $\phi_i$ is the differential phase measured for the $i$th baseline, $B_i$ is the corresponding baseline vector, and $\lambda$ is the central wavelength \citep{2003Lachaume, 2009LeBouquin}.\\
We simultaneously use the data of many baseline vectors by combining the differential phases and baseline vectors for each spectral channel and applying the Nelder-Mead simplex algorithm to find the best-fit photocenter vector using Eq. \ref{eq:photocenter}.
However, it is worth noting that Eq. \ref{eq:photocenter} is derived for marginally resolved objects.
Our AMBER data formally resolve the line-emitting region of MWC297 and therefore we note that the photocenter assumption might not provide an adequate approximation, in particular in the line center, as is discussed at the end of this Section.
The gray continuum photocenter points trace the location of the central star.\\

The derived photocenter shifts obtained from our AMBER data are shown in Fig.~\ref{fig:photocenter} (upper right panel).
Using the same method as for our CRIRES data we determine the axis of motion of our AMBER photocenters to be along a position angle of $112\pm10^\circ$, with red-shifted vectors to the south-east of the continuum and blue-shifted vectors to the north-west.
The axis of motion PA values for CRIRES ($114\pm3^\circ$) and AMBER ($112\pm10^\circ$) are consistent with each other so we adopt the CRIRES value, with smaller uncertainties, as our best estimate for the motion angle.
The PA of the axis of motion is similar to, but not fully consistent with, the PA of the disk major axis of $99.6^\circ \pm 4.8^\circ$ from our AMBER-LR observations (Sect.~\ref{Sec:2DmodelCont}).
The combined continuum and line photocenter vectors of the line wings (indicative of faster circumstellar material) are less displaced than those closer to the line center, suggesting that the closer-in circumstellar material is orbiting the star at a higher velocity than the material at large stellocentric radii.
This is consistent with the paradigm of a simple Keplerian-rotating disk or a disk wind.\\

The photocenter shifts for the lower-velocity, red-shifted channels ($\lesssim90\,$km/s) show an interesting arc-like structure where as the channels get closer to the line center they are displaced to North of the disk plane.
This does not apply for the line center (with  $|v|\lesssim10\,$km/s) in which the photocenter vectors are displaced to South of the disk plane.
A similar arc-like structure is reported by \citet{2012Kraus3} for the Herbig B[e] star \object{V921 Sco} and is speculated to result from opacity effects similar to those that we suspect are behind the arc structure seen in the CRIRES photocenter vectors (Fig.~\ref{fig:MWC297-CRIRES}).\\

The scenario outlined above can explain the arc structures seen in the line wings, but cannot account for systematic displacement in southern direction that we see in the lowest-velocity channels ($|v|\lesssim10\,$km/s; light-green points in Fig.~\ref{fig:photocenter}) and that is not observed in the CRIRES photocenters.
We observe this effect only in the three line channels in the very center of the line, where the line-emitting region is most extended (see Fig.~\ref{fig:linevis}, right panel).
In these channels, the line-emitting region is strongly resolved (with visibilities of a few per cent) and the approximation in Eq.~\ref{eq:photocenter} is likely not applicable anymore. Therefore, we believe that the systematic displacement observed in these channels is an artifact, indicating that the geometry of the line-emitting region is too resolved to be reasonably represented by a simple photocenter displacement.
This conclusion is also supported by the lower panel shown in Fig.~\ref{fig:photocenter}, where we compare the differential phases that correspond to the photocenter displacement model (black line) with the actual measurements and where we observe significant residuals in the line center.

\section{Velocity-resolved image reconstruction}
\label{Sec:image-reconstruction}


\begin{figure*}
\centering
\includegraphics[width=18.4cm]{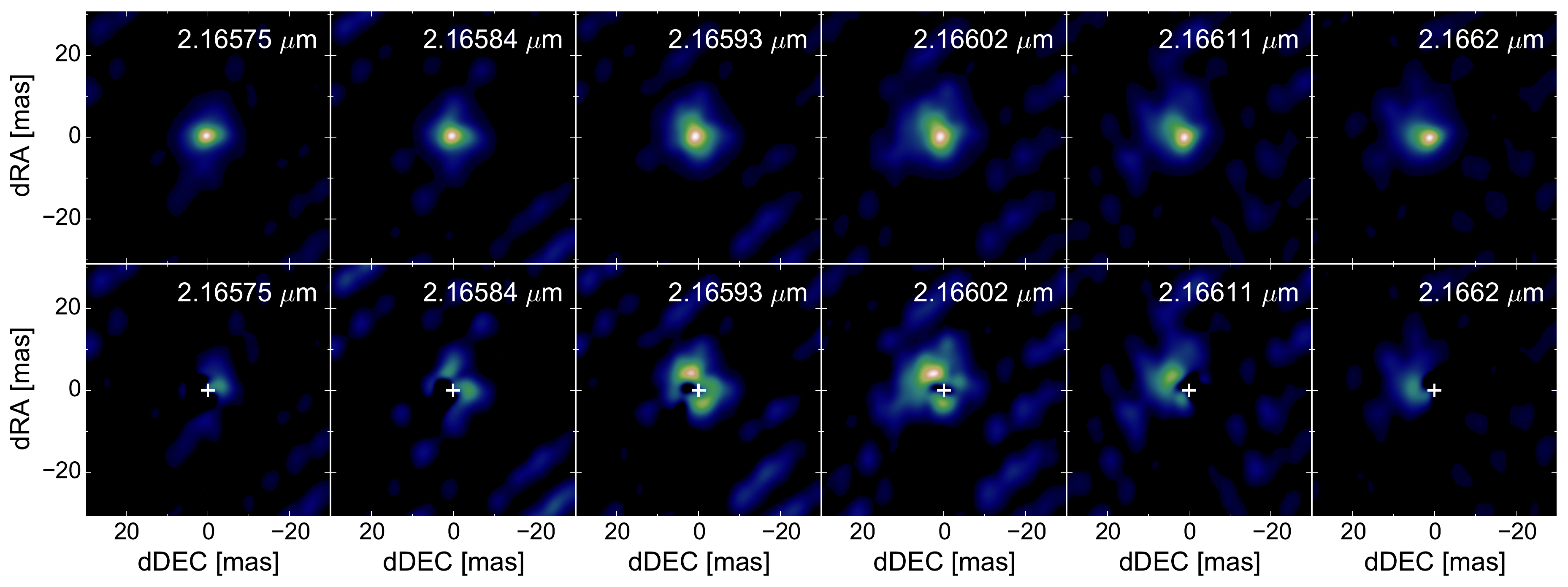}
\caption{Velocity-resolved aperture synthesis images of MWC297 calculated by applying the \emph{IRBis} software to our AMBER HR-K data. The raw images are shown in the upper panels and the continuum-corrected images are shown in the lower panels. The corresponding wavelength for each spectral channel is shown in the upper right corner of each image and the location of the continuum center is marked as a white cross in the continuum-corrected images.}
\label{fig:imagefigs}
\end{figure*}

\begin{figure*}
\centering
\includegraphics[width=18.4cm]{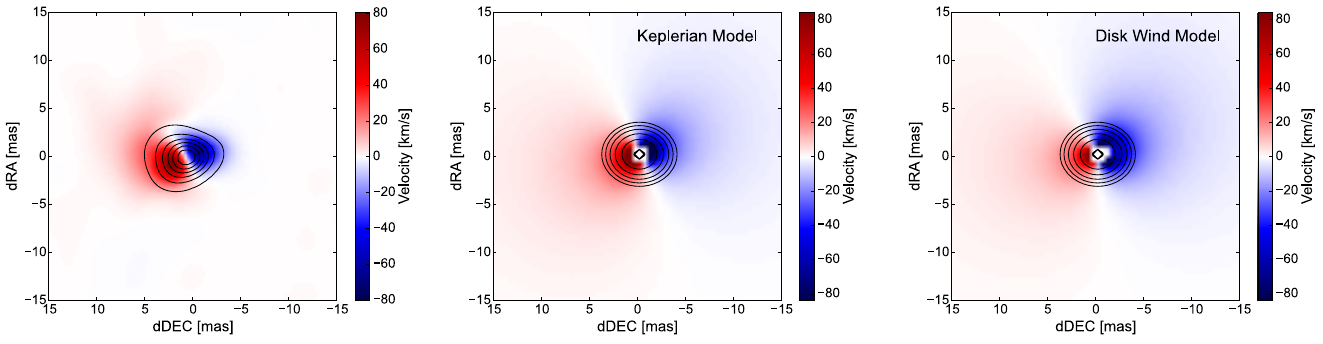}
\caption{Left panel: First moment map for the Br$\gamma$ line computed from our MWC297 AMBER images (shown in Fig.~\ref{fig:imagefigs}, upper panel). Black contours show the continuum emission region. Middle panel: First moment map constructed from the frames of our Keplerian disk model. Right panel: First moment map constructed from the frames of our disk wind model.}
\label{fig:image-moment}
\end{figure*}

Interferometric imaging is a useful way to gain a model-independent representation of the brightness distribution.
Spectrally dispersed interferometry enables us to obtain such images in multiple spectral channels, revealing wavelength-dependent differences due to gas kinematics.
Velocity-resolved images in infrared line tracers such as Br$\gamma$ have already been obtained for a supergiant A[e] star \citep{2011Millour} and a Luminous Blue Variable \citep{2016Weigelt, 2017GRAVITY}.
Here, we apply this method to obtain the first velocity-resolved images for a young star at infrared wavelengths.
We use the \emph{IRBis} method for image reconstruction \citep{2014Hofmann} in order to reconstruct images for each channel across the Br$\gamma$ line, using the differential-phase method outlined in \citet{2011Millour} and \citet[][Sect.~2]{2016Weigelt}.
The obtained images were convolved with the point-spread function of a single-dish telescope with a diameter which is twice the length of the longest baseline in our $uv$-plane.
We see that the brightness distribution clearly changes  with wavelength in our image cube, reflecting the strong signals detected in the wavelength-differential visibilities and phases across the Br$\gamma$ line.
Comparing the images at different velocity channels (Fig.~\ref{fig:imagefigs}) we see that the line emission is clearly more extended in the two central channels ($2.16593\mu$m and $2.16602\mu$m) than in the high-velocity channels ($2.16575\mu$m and $2.1662\mu$m).
In order to display the geometry of the line-emitting region with better contrast, we normalize the flux level of each image equal to the corresponding flux from the AMBER spectrum and then subtract the continuum intensity distribution, with the results shown in Fig.~\ref{fig:imagefigs} (lower panel).
Examining the continuum-subtracted images we can see that the morphology of the line-emitting region changes significantly across the Br$\gamma$ line.
In the highest-velocity channels (wavelengths of $2.16575\mu$m and $2.1662\mu$m, $25 \lesssim |v| \lesssim 37$\,km/s) the emission is displaced to the West (right) in the blue-shifted line channels and to the East (left) for the red-shifted channel.
In the channels with intermediate velocity ($2.16584\mu$m and $2.16611\mu$m, corresponding to $12 \lesssim |v| \lesssim 25$\,km/s) the material is more extended, but still displaced in the same direction as in the lower-velocity channels.
Additionally the shape of the brightness distribution changes from a single-lobed structure at the highest velocities ($2.16575\mu$m and $2.1662\mu$m) to a double-lobed structure with two brightness peaks at intermediate velocities, one lobe further to the North and one further to the South (seen most clearly in the $2.16584\mu$m channel).
The line-emitting region is most extended in the line center and most compact in the line wings.
At these lowest-velocity channels ($2.16593\mu$m and $2.16602\mu$m, $|v| \lesssim 12$\,km/s) the brightness distribution shows an extended ring-like structure, with a more pronounced northern lobe than the southern side.
This can be viewed as a continuation of the previously discussed double-lobed structure where the brightness distribution is now more extended to the North and South than in the East-West direction.\\

To further study the kinematics, we construct a first moment map from our image cube, where we compute the first moment coordinate value $M_1$ with 
\begin{equation}
M_1 = \frac{\sum{I_i v_i}}{v_\text{norm}},
\label{eq:1stmoment}
\end{equation}
where $I_i$ is the pixel value, $v_i$ is the velocity of the image frame relative to the line center and $v_\text{norm}$ is a factor that normalizes the maximum absolute value of $M_1$ to the total width of the image cube in velocity space.
The resulting 2D velocity field is shown in Fig.~\ref{fig:image-moment} and supports what we see in the continuum-subtracted images, with red-shifted material displaced to the East of the star and blue-shifted material displaced to the West.
By finding and measuring the angle between the respective peaks of the blue- and red-shifted emission lobes we can estimate the "axis of motion" (identical to the disk axis in the case of a Keplerian-rotating disk) of $114^\circ$, a very good match to the photocenter displacement vectors of both AMBER ($112\pm 10$) and CRIRES ($114\pm 3$).


\section{Kinematic modeling}
\label{Sec:kinematicmodel}


In order to obtain quantitative constraints on the gas velocity field around MWC297 we fit a kinematic model to our AMBER observables exploring different scenarios for the origin of Br$\gamma$ emission, namely a simple Keplerian-rotating disk and a rotating disk with an outflowing velocity component corresponding to a simplified disk-wind scenario.
We use a kinematic modeling code that has already been used for modeling spectro-interferometric observations on a range of evolved \citep{2007Weigelt,2012Kraus1} and young \citep{2012Kraus2} objects.
Based on an analytic description of 3D velocity fields and the radial brightness distribution, this code allows us to compute synthetic spectra and synthetic images for different velocity channels in a spectral line.
This type of modeling is a useful tool to explore how different disk morphologies and velocity fields fit the observed visibilities and differential phases.
We model multiple physical scenarios that are described in the following two subsections, including a Keplerian rotation field in a geometrically thin disk and a velocity field similar to a magneto-centrifugally driven disk wind \citep{1982Blandford}.\\

\begin{figure}
\centering
\includegraphics[width=\columnwidth]{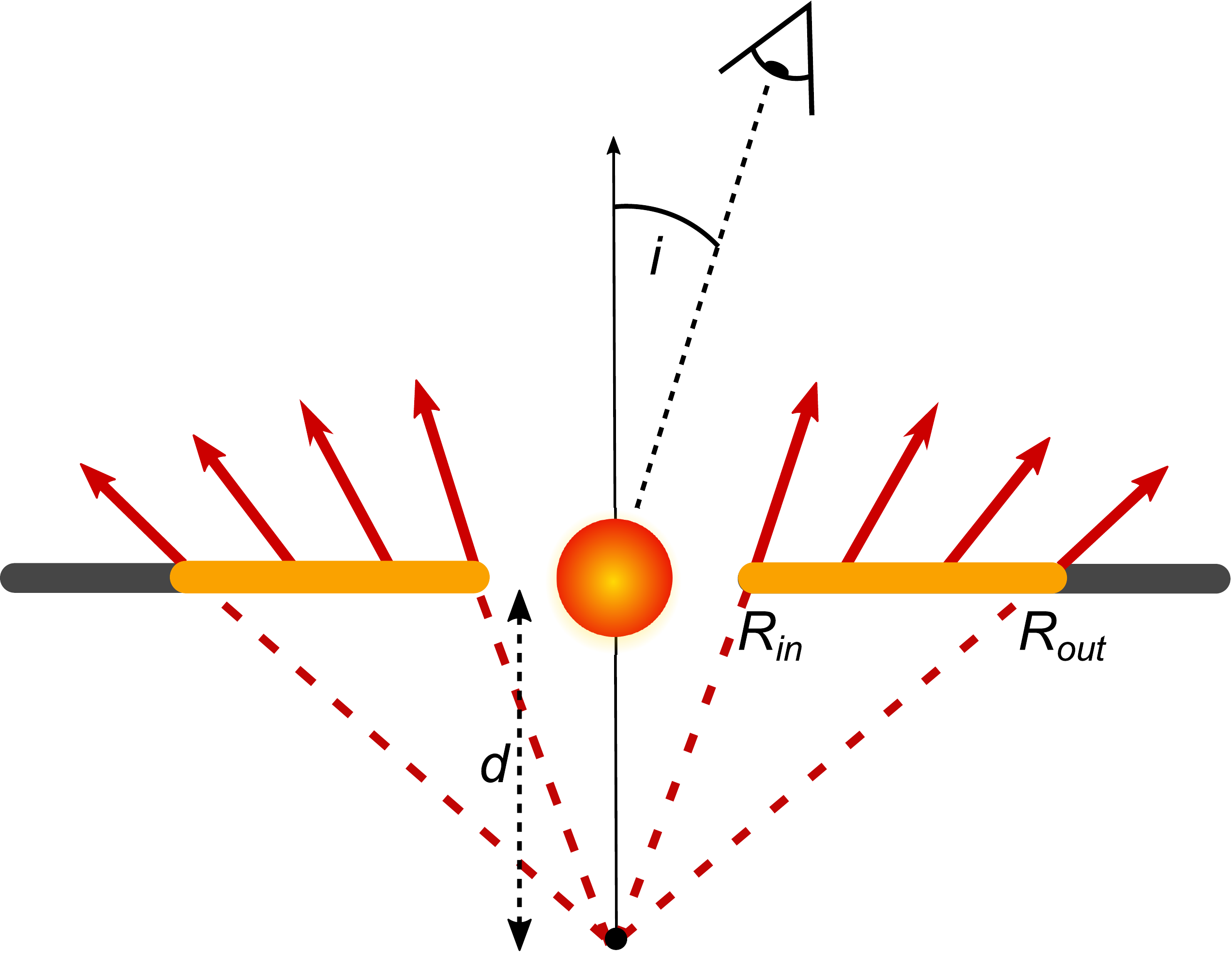}
\caption{In our kinematic model the Br$\gamma$ emission is emitted from a region of the disk's surface between $R_{in}$ and $R_{out}$ (shown in yellow) and is viewed at an inclination angle $i$. The kinematics of the line-emitting region include a toroidal component (analogous to $v_{k}$) and poloidal component, shown as red vectors from an imaginary source displaced at a distance $d$ from the center of the system.}
\label{fig:schematic}
\end{figure}

For both scenarios, we compute synthetic images that include contributions from a continuum-emitting component and the line-emitting component.
The continuum is modeled with an inclined Gaussian brightness distribution, using the FWHM size $D$, inclination $i$ and system axis position angle $\theta$ determined in Sect.~\ref{Sec:2Dmodel}.
The line-emitting region is modeled to extend from an inner radius $R_{in}$ to an outer radius $R_{out}$, while the radial brightness distribution follows the power-law $\propto r^\beta$, where $r$ is the radius and $\beta$ is a free parameter (see Fig.~\ref{fig:schematic} for a schematic of the model setup).
To avoid unphysical sharp edges in our model images (which might introduce artifacts), we implement the outer truncation using a Fermi-type smoothing function \citep[see][]{2008Krausb}.
To model the arc structures we see in the photocenter displacements, we introduce an opacity gradient (multiplied by an arbitrary factor $\alpha_o$) which darkens the more distant parts of the disk.
We adopt the stellar parameters (mass $M_{\star}$ and distance $d$) quoted in Sect.~\ref{Sec:intro}.
From the resulting 3D image cube we compute wavelength-differential visibilities and phases at the $uv$-coordinates covered by our data.

\begin{figure*}
\centering
\includegraphics[width=18.4cm]{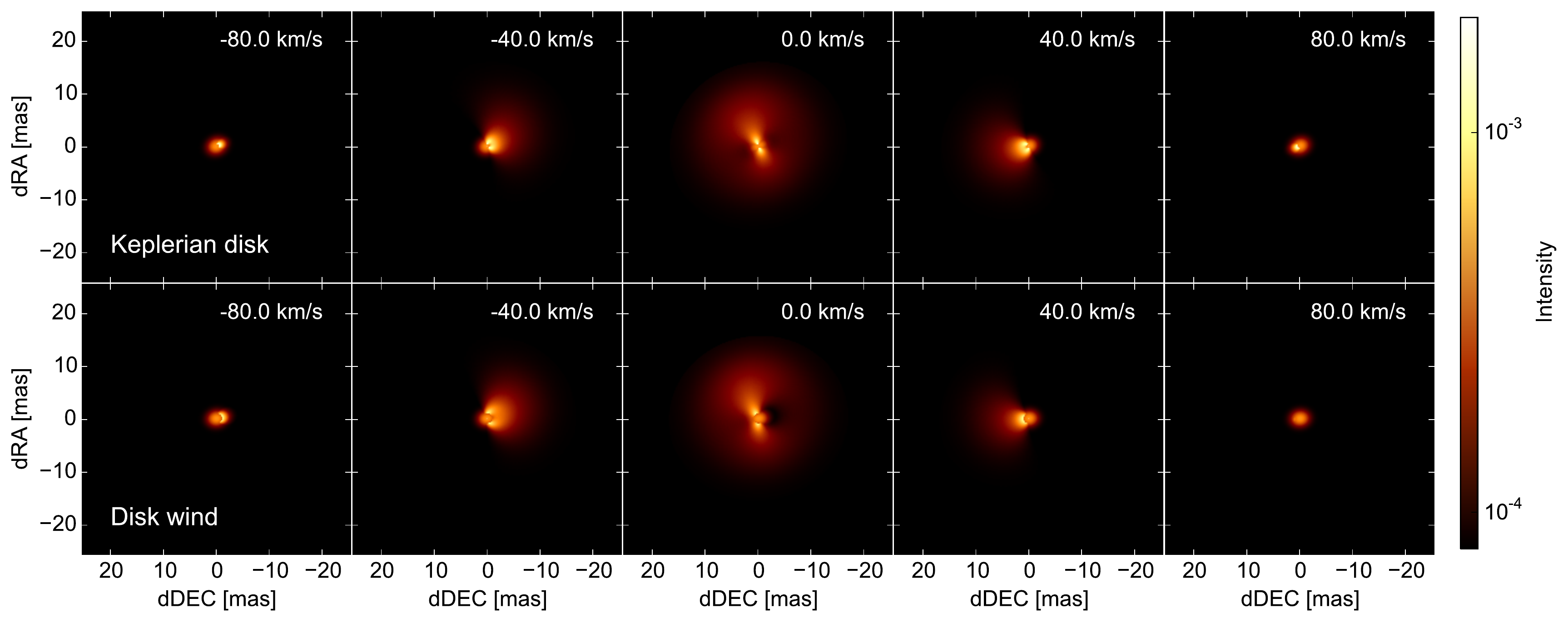}
\caption{Upper panel: Synthetic model images calculated for our Keplerian model for five spectral channels. We use an arbitrary logarithmic color scale to show the brightness distribution. Lower panel: Synthetic model images calculated for our disk wind model for five spectral channels. We use an arbitrary logarithmic color scale to show the brightness distribution.}
\label{fig:simmap-split}
\end{figure*}

\begin{table}
\caption{Ranges and best fit values for our kinematic model parameters. The reduced $\chi^2$ values for each of our best fit kinematic models are also shown.}
\label{tab:modelparameters}
\centering
\vspace{0.1cm}
\begin{tabular}{c c c c}
\hline \hline
Parameter & Range & Keplerian & Disk wind\\
\hline
$R_{in}$ & $0$ - $20$ mas & $1.7{\pm0.5}$ mas & $1.9{\pm0.5}$ mas\\
$R_{out}$ & $0$ - $100$ mas & $45{\pm4}$ mas & $45{\pm4}$ mas\\
$\theta$ & $0^\circ$ - $180^\circ$ & $202.3{\pm1.9^\circ}$ & $192.1{\pm2.1^\circ}$ \\
$i$ & $0^\circ$ - $90^\circ$ & $21.8{\pm3.5^\circ}$ & $23.1{\pm3.4^\circ}$ \\
$\beta$ & $-2.0$ - $0.0$ & $-0.99{\pm0.08}$ & $-0.99{\pm0.08}$ \\
$v_z$ & $0.0$ - $2.0 \times v_k$ & - & $0.14{\pm0.02}\times v_k$\\
$d$ & $0$ - $45$ mas & - & $1.1{\pm0.4}$ mas \\
\hline
$\chi^2_{r,V}$ & - & $0.45$ & $0.39$ \\
$\chi^2_{r,\phi}$ & - & $1.19$ & $1.11$ \\
$\chi^2_{r}$ & - & $1.64$ & $1.50$ \\
\hline \hline
\end{tabular}
\end{table}

\begin{figure}
\centering
\includegraphics[width=70mm]{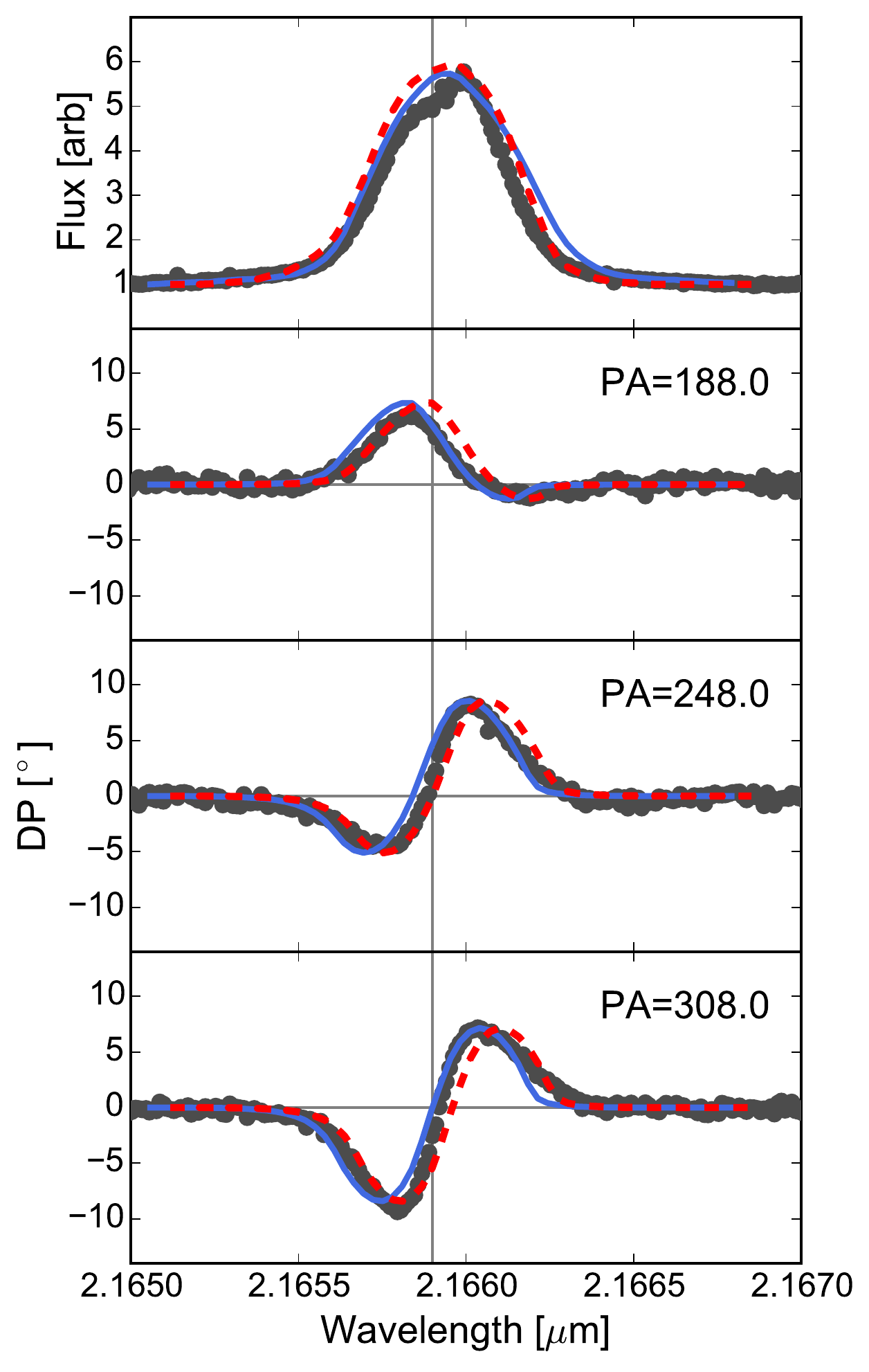}
\caption{Synthetic spectra (top row) and differential phases (lower rows) from our Keplerian (blue solid line) and disk wind (red dashed line) models plotted over our observed CRIRES data (dark gray points).}
\label{fig:CRIRESmodel}
\end{figure}

\subsection{Keplerian disk model}

We first consider a simple Keplerian disk model, where the Br$\gamma$-emitting region extends from an inner radius $R_{in}$ to an outer radius $R_{out}$. 
In the Keplerian rotation model, the velocity field $v(r)$ is defined as $v(r) = \sqrt{\frac{GM}{r}}$, where $G$ is the gravitational constant (we subsequently refer to this velocity field as $v_k$).
We systematically sample the parameter space and compute the reduced $\chi^2$ between the measured and model visibilities and phases: $\chi^2_r = \chi^2_{r,V} + \chi^2_{r,\phi} $.
The parameter uncertainties are calculated by fitting the $\chi^2$ surface near the minimum.
The model parameter values and uncertainties corresponding to the best-fit model are listed in Table~\ref{tab:modelparameters} and the corresponding model visibilities/phases/spectra are shown in Figs.~\ref{fig:simmap-models} and \ref{fig:simmap-spec} (solid blue lines).
The model is able to reproduce the visibility drop across the line well for the majority of intermediate to long baselines, but for the shortest (< 42m) baselines the model visibilities are lower than the observed quantities across the line.
The sign and shape of the differential phase measurements are reproduced for the majority of baselines, with some significant residuals visible for the red-shifted part of the baselines with PAs between $-100^\circ$ and $-96^\circ$.
The single-peaked spectral line is modeled well with this scenario.\\

After fitting the AMBER data we also determine how the photocenter displacements from our best fit model images compare to our observed CRIRES spectro-astrometry data.
Figure \ref{fig:CRIRESmodel} shows that the Keplerian model is a good fit to the CRIRES data, able to recreate the features of the DPs and the single peaked shape of the emission line.\\
Despite this model providing a generally good fit to the AMBER HR-K interferometry data, we find that the disk PA in the best-fit model does not match the disk orientation of the continuum-emitting disk, as modeled in Sect. \ref{Sec:2Dmodel}.
We find a $\theta$ value of $202.3^\circ$ for our Keplerian model which is a perfect match to the axis of motion measured in Sect. \ref{Sec:photocenter} but shows a ${\sim}3\sigma$ deviation from our AMBER LR measurement of the continuum PA.
This calls the validity of the model into question, as it is an intrinsic property of a Keplerian disk model that the PA of the line displacement must be aligned with the major axis of the dust continuum disk.

\subsection{Disk wind model}
\label{subsec:dwmodel}

Given the remaining residuals in the Keplerian disk model as well as the significant deviations between the continuum disk orientation and the Br$\gamma$-line photocenters (Sect. \ref{Sec:photocenter}), we investigate in this section a more complex gas velocity field, such as a velocity field similar to a disk wind.
For this purpose, we add to our Keplerian disk model a velocity component that points out of the disk plane, as predicted by theoretical accretion-driven MHD disk wind models \citep[e.g.,][]{1982Blandford, 1999Goodson, 2006Ferreira}.
We assume that the line emission originates from the hot gas located near the disk, oriented relative to the observer with the disk inclination $i$ and major axis PA $\theta$, with both quantities treated as free parameters.
In this model, the gas velocity is modeled as a superposition of the toroidal velocity component $v_\phi$ (identical to the Keplerian velocity $v_k$ used in our previous model and a poloidal component representing the motion along the magnetic field lines along which material is accelerated and lifted off the disk surface.
To parameterize this poloidal velocity component we adopt a geometry analogous to the one discussed in \citet{2006Kurosawa}, with wind streamlines drawn from an imaginary point on the opposite site of the disk plane and located at a distance $d$ (Fig.~\ref{fig:schematic}).
The velocity along this poloidal vector is treated as a free parameter and scales as a function of the Keplerian velocity $v_k$.\\

\begin{figure*}
\centering
\includegraphics[width=18.4cm]{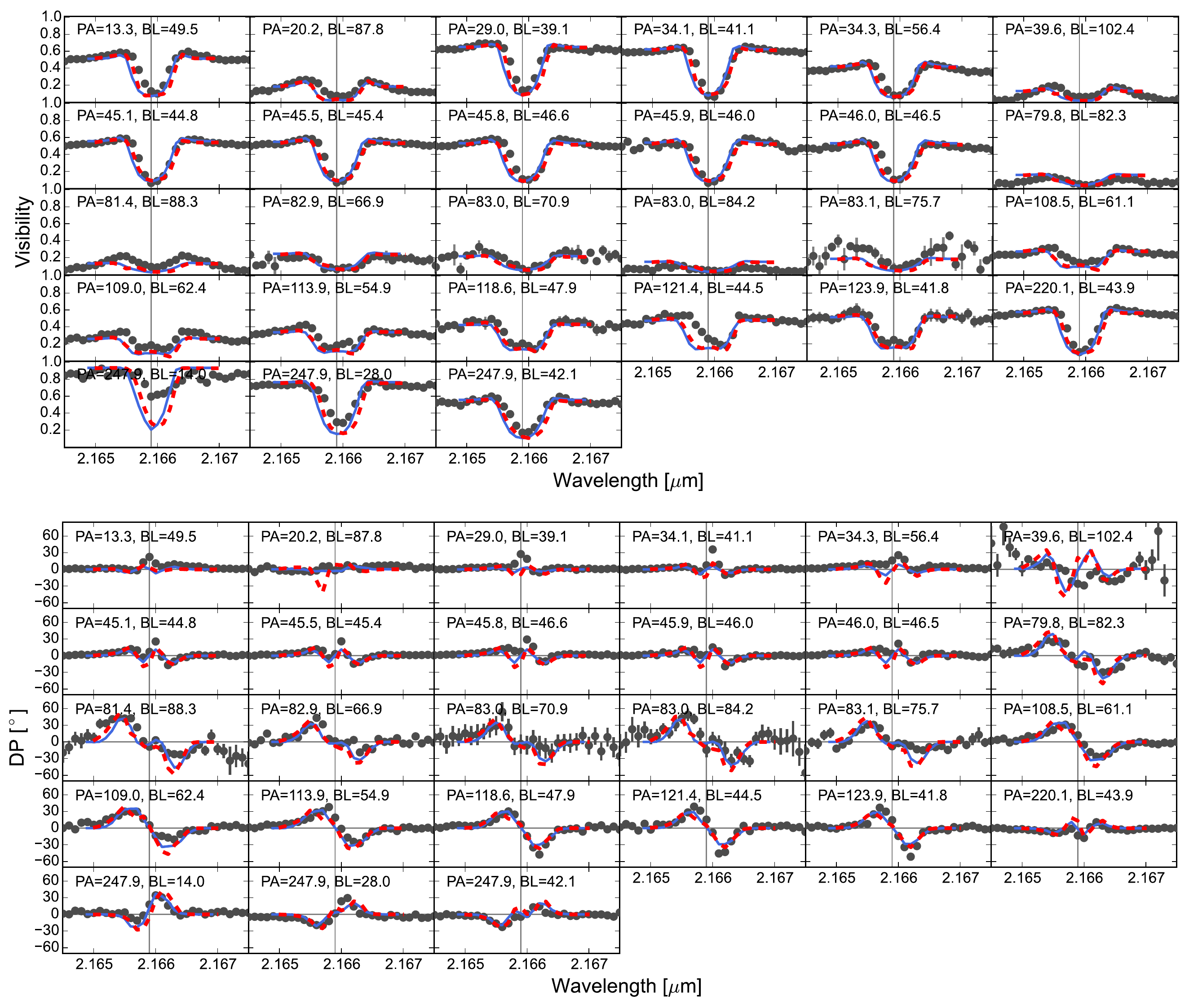}
\caption{
Visibilities (upper panel) and differential phases (lower panel) calculated from our kinematic models and compared to our observed data.
The solid blue line represents the simple Keplerian disk model and the dashed red line represents our simple disk wind model.
Data points are gray.
The frame with PA=39.6, BL=102.4 is offset by $180^{\circ}$ in order to see the differential phase changes at the Br$\gamma$ line center.
}
\label{fig:simmap-models}
\end{figure*}

\begin{figure}
\centering
\includegraphics[width=\columnwidth]{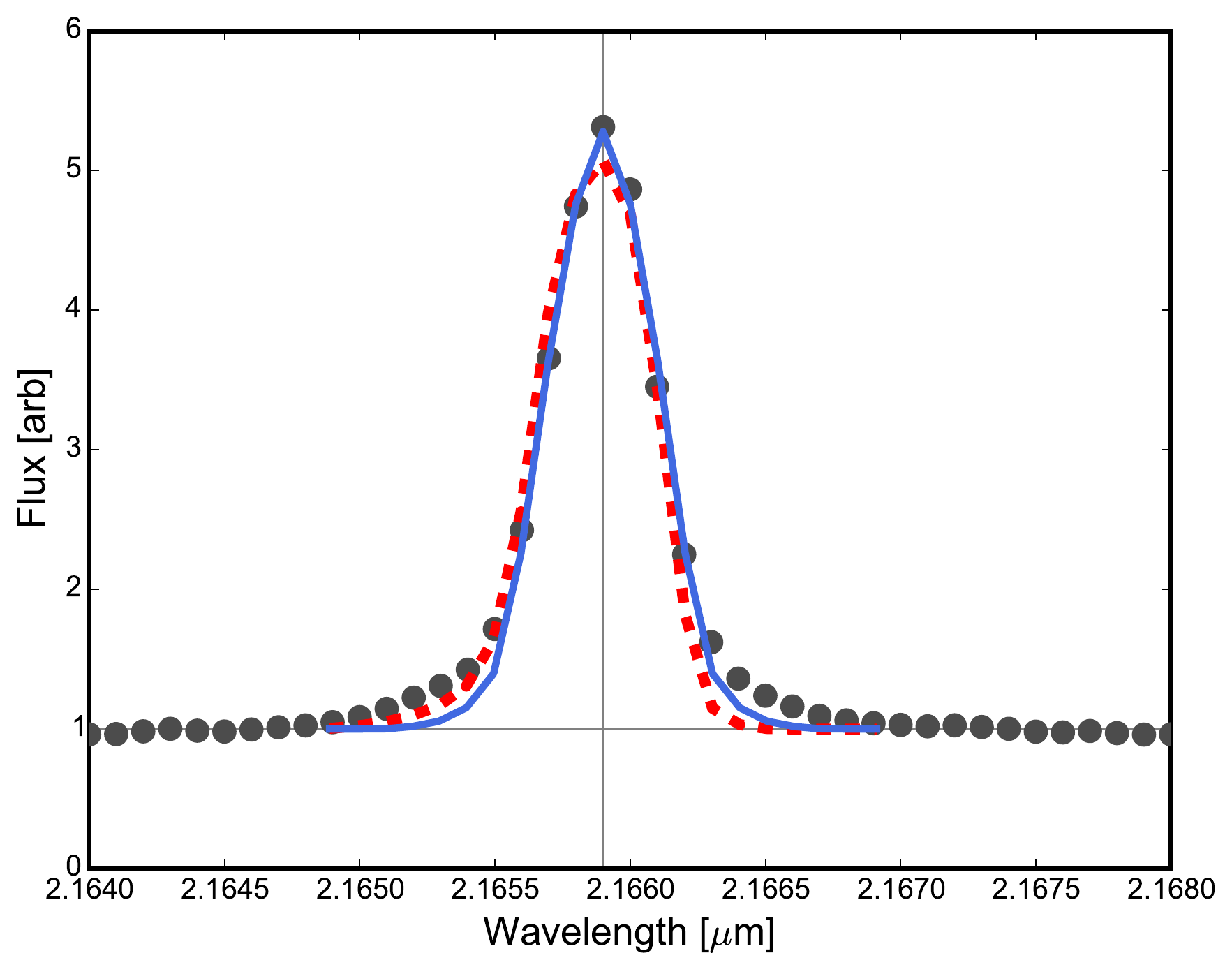}
\caption{
Spectrum calculated from our kinematic model compared with the Br$\gamma$ spectrum from our AMBER data set.
The solid blue line represents the Keplerian model and the dashed red line represents the simple disk wind model.
Data points from AMBER are gray.
}
\label{fig:simmap-spec}
\end{figure}

The synthetic images from our simple disk wind model are shown in Fig. \ref{fig:simmap-split} (lower panel) and its comparison to our observed data can be seen in Figs. \ref{fig:CRIRESmodel} (CRIRES) and \ref{fig:simmap-models} (AMBER).
Details of the best fit values for all free parameters for both the disk wind and Keplerian models are listed in Table \ref{tab:modelparameters}.
The angular size of both the line and continuum-emitting regions for the disk wind model are nearly identical to those of the Keplerian model, causing the visibilities (which predominantly trace the object's 2D geometry) for each model to be very similar.
The differences in the kinematic paradigm cause changes in the differential phase, however, due to the small out-of-plane velocity, these changes from the Keplerian model are subtle and allow the model to much better fit some of the differential phase measurements.
We see no major deviations in the shape of the line profile, only a small shift of the line caused by the out-of-plane component (see Fig.~\ref{fig:simmap-spec}).
The new model including the out-of-plane velocity component results in a much better fit to the PA of the continuum disk than the fit of the Keplerian disk model.\\

\begin{figure*}
\centering
\includegraphics[width=18.4cm]{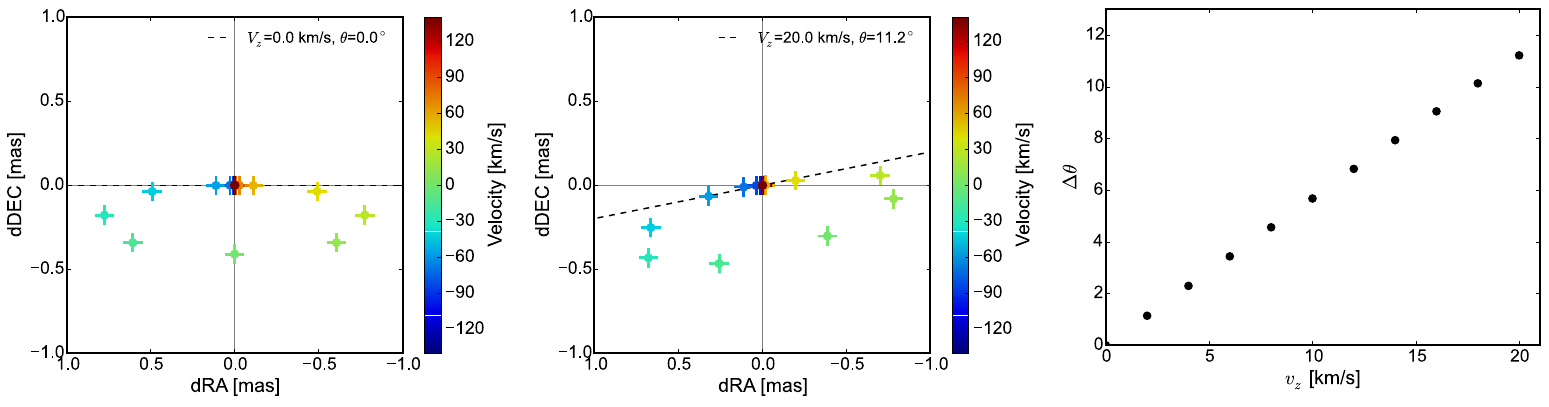}
\caption{
Left: Model photocenter displacements for our Keplerian disk model with the disk major axis along a PA of $90^\circ$. We impose an opacity field in order to shift the central channels away from the continuum so that they do not overlap. Center: Model photocenter displacements for a similar model with an out of plane velocity of $20$\,km/s. We see that the perceived PA of the system axis is offset by a value of $11.2^\circ$ and that the red-shifted vectors are less displaced from the continuum than their blue-shifted counterparts. Right: A plot of the out-of-plane velocity ($v_z$) plotted against system axis change ($\Delta\theta$) for our grid of models showing that the PA offset varies linearly with $v_z$.
}
\label{fig:model-angles}
\end{figure*}

Throughout our kinematic modeling process we noted that adding an out-of-plane velocity component, such as is present in a disk wind, changes the perceived rotation angle traced by the Br$\gamma$ photocenter vectors.
To explore this phenomenon further we computed a grid of similar disk wind models, changing only the out-of-plane velocity for each different model, and calculated their photocenter displacement profiles from the raw model fit images.
Using a simple linear fit, we determine the position angle along which the photocenter vectors are predominantly displaced (see Fig. \ref{fig:model-angles}, left and center panels) and find that the position angle offset increases linearly with increasing out-of-plane velocity (Fig. \ref{fig:model-angles}, right panel).
We also find that as the wind velocity increases, the red-shifted photocenters become less displaced from the center, with the blue-shifted vectors becoming more displaced.
We therefore propose that measuring the displacement of the perceived axis of motion from the known disk major axis is a powerful diagnostic for the presence of out-of-plane velocity components.
Further study into the physical interpretation of the PA shift, how it depends on the different parameters of the disk wind model, and an investigation of the phenomenon using other kinematic disk-wind models will be the subject of a future study.

\section{Discussion}

    By measuring the photocenter displacement vectors from both spectro-interferometry and spectro-astrometry, we gain a model-independent view of the kinematics traced by the Br$\gamma$ line.
    A remarkable trait of the photocenter patterns that is present in the AMBER and CRIRES data independently, is the looping arc structures.
    An example of a similar arc structure observed with IR spectro-astrometry can be found in the presentation of \object{V921 Sco} CRIRES data by \citet{2012Kraus3}, although the displacement of the central channels away from the continuum was not as extreme as the case of MWC297 that we present in this paper.
    The ratio of the major/minor axes of the loop is dependent on the opacity parameter $\alpha_o$ and the inclination of the disk (minor axis becomes larger for more face-on objects).
    We account for this pattern using an opacity field that obscures the furthest reaches of the disk, however, the origin of this opacity is not described in the model and there are a number of different scenarios that could be causing this.\\
    In their recent work using SMA to study a massive young stellar object, \citet{2016Ilee} observed similar looping arc-like structures in the centroid shifts, traced by multiple spectral lines, which were the result of displacement of the central channel centroids away from the disk plane.
    Their system is seen closer to edge-on ($i{\sim}55^\circ$) and they interpret the arc-shaped displacement as being caused by the flaring outer regions of the disk.    
    It is unlikely that this scenario applies to MWC297 as the disk is viewed almost pole-on ($i {\sim}20^\circ$) and any disk flaring will have only a small effect on the opacity so close to the center of the disk.
    Therefore, other interpretations such as optically thick dust above the disk plane might be more applicable in the case of MWC297.
    \citet{2012Bans} theorized that dust can exist above the disk plane. 
    Theoretical explanations for this include the ejection of dust in a disk wind \citet{2012Bans} and a dusty magnetically supported atmosphere \citet{2014Turner}.
    
    At the longest baseline of our AMBER data set we see a $180^\circ$ shift in the differential phase across the Br$\gamma$ line (PA=$39.6^\circ$, BL=$102.4$m).
    In Figs. \ref{fig:photocenter} and \ref{fig:simmap-models} the range of the frame is shifted in order to show the differential pattern of the phase and the comparison to the models.
    This phase jump at long baselines is indicative of the visibility moving from the first to the second lobe, caused primarily by sharp edges in the observed brightness profile.
    The fact that we only see this phase jump in the Br$\gamma$ line and at the longest baselines indicates a detection of the sharp inner edge of the line-emitting region.\\
    
    Using the \emph{IRBis} method we reconstruct a series of velocity-resolved images across the Br$\gamma$ line from our AMBER HR-K data set, the first such images of a young star achieved with optical interferometry (Sect.~\ref{Sec:image-reconstruction}).
    We retrieved images for six spectral channels within the line and computed continuum-subtracted images (Fig.~\ref{fig:imagefigs}) as well as a first-moment map (Fig.~\ref{fig:image-moment}) that reveals the 2D velocity field, with the blue- and red-shifted lobes displaced along a position angle of $114^\circ$. This is in agreement with the photocenter displacement vectors derived from both our AMBER and CRIRES data sets (Sect.~\ref{Sec:photocenter}).\\
    The wavelength-dependent brightness distribution in the images
closely resembles the synthetic images of both our Keplerian and disk wind models, both in the morphology of the line-emitting gas and the perceived "axis of motion".
    At intermediate velocities ($2.16584\mu$m and $2.16611\mu$m), we see a double-lobed structure, both in the reconstructed images and the model images ($\pm40$km/s), although it is not possible to distinguish between the Keplerian and disk wind scenario using the images alone.\\
    
    To provide a comparison between our kinematic model and the image reconstructions we constructed first moment maps from our model frames using the same code as was used for the images in Sect.~\ref{Sec:image-reconstruction}, with the results shown in the middle and right-hand frames of Fig.~\ref{fig:image-moment}.
        Both the Keplerian and the disk wind models show the displacement of blue- and red-shifted material along a similar PA to what was seen in the photocenter shifts of both AMBER (Fig.~\ref{fig:photocenter}) and CRIRES (Fig.~\ref{fig:MWC297-CRIRES}), as well as along a similar PA as seen in the moment map from the images (Fig.~\ref{fig:image-moment}, left panel).
        The most noticeable difference in the moment maps of the two model scenarios is that the blue-shifted lobe of the disk wind model traces a larger area than the corresponding lobe in the Keplerian model.
        Whilst there are some slight differences between the different models' moment maps, both moment maps are a good match to the results of the image reconstruction (Fig.~\ref{fig:image-moment}, left panel) and we find that comparing the moment maps is not a useful way to distinguish between models.
        The calculation of the moment map from our model frames using the same code as was used for the reconstructed images creates a small square shaped artifact at the center of the moment map caused by the rotation of the inner edge of the line-emitting region.\\
        
    Disks demonstrating Keplerian-like motion have also been observed around various other young stellar objects (\object{V921\,Sco}, \citealt{2012Kraus3}; \object{HD\,100546}, \citealt{2015Mendigutia}; \object{HD\,58647}, \citealt{2016Kurosawa}).
    A significant difference between these cases and our observations of MWC297 can been seen in the relative geometries of the continuum and line-emitting regions.
    For the objects mentioned above, particularly \object{V921 Sco} and \object{HD\,100546}, the Br$\gamma$-line emission is more compact than the K-band continuum emitting ring, usually tracing gas inside or near the dust inner rim (in contrast to MWC297).
    Our analysis and modeling of the wavelength-dependent visibilities shows that the line-emitting gas is located in a very extended region relative to the continuum, though this could be due to the extremely compact K-band continuum size.
    The compact nature of the continuum has already been discussed by many other authors \citep{2004Eisner,2005Monnier,2007Malbet,2008Kraus,2008Acke,2011Weigelt} but the physical interpretation for this compact continuum size is still a subject of discussion.
    This problem of undersized Herbig Be stars is not exclusive to MWC297, and was also found in several other high-luminosity objects in the size-luminosity study of \citet{2002Monnier}.
    Several general explanations as to why these objects are so compact have been put forward, including (a) optically thick gas emission from a compact viscously heated accretion disk \citep{2008Kraus}, (b) an inner gaseous component that shields stellar radiation to let dust survive closer to the central star \citep{2004Eisner,2002Monnier}, and (c) additional emitting components such as highly refractory grains \citep{2010Benisty}.\\
    
    Despite the extremely compact K-band continuum emission, the relative extension of the Br$\gamma$ emission raises the question of how the line-emitting gas can be heated to the temperatures required to emit the Br$\gamma$ line \citep[i.e., $8000-10,000$K,][]{2006Kurosawa}.
    The compact nature of the K-band continuum requires optically thick material at small circumstellar radii, and therefore it is unlikely that the more extended line-emitting gas could be heated by radiation from the central star.
    In their previous work on this object, \citet{2011Weigelt} discussed the relationship between the continuum and the line-emitting materials and favoured the explanation that the compact continuum emission is caused by a mixture or warm dust and refractory grains which possibly play an important role in the formation of a disk wind.
    Material ejected in a disk wind can be rapidly heated by ambipolar diffusion up to temperatures sufficient to emit in the Br$\gamma$ line, and these temperatures can also be exceeded in the disk midplanes of rapidly accreting objects.\\
    We find that our best-fit Keplerian kinematic model shows a 3$\sigma$ deviation from the continuum disk PA, a difference that we account for by employing a disk wind model which results in a similar PA to the continuum.\\
    
    Using the results from our photocenter analysis allows us to place tight constraints on the position angle of the rotation that we observe in the Br$\gamma$ line.
    We can see from our observations in Figs. \ref{fig:MWC297-CRIRES} and \ref{fig:photocenter} that our observed CRIRES and AMBER photocenters are consistent with one another, with axes of motion of $112\pm10^\circ$ (AMBER, Sect.~\ref{subsec:amberp}) and $114\pm3^\circ$ (CRIRES, Sect.~\ref{subsec:criresp}), which results in our best estimate for the Br$\gamma$-line position angle of $114\pm3^{\circ}$.
    This angle differs significantly from the major axis of the dust disk intensity distribution, as determined from near-infrared interferometry in the H- and K-band continuum.
    Measurements of the continuum geometry from PIONIER \citep{2017Lazareff} and AMBER estimate the position angle of the disk major axis to $103.7\pm1.7^\circ$ and $99.6\pm4.8^\circ$, which are consistent with each other, but both differ from our measurements of the axis of motion by ${\sim}10^\circ$ (detailed in Sect. \ref{Sec:photocenter}, Figs. \ref{fig:MWC297-CRIRES} and \ref{fig:photocenter}).
    Whilst modeling the kinematics of the disk wind (Sect.~\ref{subsec:dwmodel}) we noted that the apparent PA of the axis of motion became displaced from the continuum disk axis as the out-of-plane velocity increased (shown in Fig. \ref{fig:model-angles}).
    We find that the observed PA offset of ${\sim}10^\circ$ can be explained by an out-of-plane velocity between $15$ and $20$ km/s.
    When exploring the disk wind model parameter space we found that the best fit model is consistent with the continuum geometry as measured by PIONIER and AMBER as well as providing a good fit to the visiblity and differential phase measurements from our AMBER HR-K data.
    As a possible alternative explanation for the percieved PA difference we considered a warped-disk scenario, where the orientation of the disk changes as a function of radius.
    In this case, the continuum emission would trace a closer-in region of the disk than the Br-$\gamma$ line emission and the observed PA difference would trace the radial differences in disk orientation instead of kinematical effects.
    However, as the line emitting region is only approximately two times larger than the continuum disk, the warping would have to be very extreme to produce  a PA difference as large as the $10^\circ$ that we see.
        Additionally, we see no evidence of a PA difference between the H-band PIONIER observations, which find a PA of $103.7\pm1.7^\circ$ \citep{2017Lazareff}, and our AMBER K-band measurements with a PA of $99.6\pm4.8^\circ$.
        The consistency of these two observations, which trace regions of the disk with different radial extension, suggests that the MWC297 disk is not significantly warped but more measurements of the disk PA at larger radial extension could further clarify this.\\
    
    We test the fit of our disk wind model to the interferometric observations of \object{MWC297} and examine how adding an out-of-plane velocity component affects the synthetic channel maps and Br$\gamma$ line profiles.
    It is easy to spot the main differences between our Keplerian model and disk wind model when we compare their synthetic model images (Fig. \ref{fig:simmap-split}).
    Keplerian rotation produces symmetric brightness profiles in the channel maps, while the disk wind model permits also asymmetric profiles.
    In our disk wind model we can see a clear shift away from the symmetric velocity field of the Keplerian paradigm, with more flux visible in the blue- than in the red-shifted
channels and an asymmetric brightness distribution in the central-wavelength channel image.
    We calculate the blue-shift caused by the out-of-plane velocity in the model and apply a corresponding wavelength correction to correct for the small wavelength changes we see in the disk-wind model.
    We find that the differences in the synthetic data between our Keplerian model and the disk wind model are subtle, and manifest themselves mainly as  differences in the differential phase.
    For each scenario we compare the observed data with the synthetic data from each of our models and calculate the reduced $\chi^2$, finding a marginal difference between the two models with the Keplerian model having a larger $\chi^2$ ($1.64$) than the disk-wind model ($1.50$).
    In particular, the disk-wind model is able to reproduce small phase jumps in the center of the Br$\gamma$ line that are seen for baselines with position angles between $80^\circ$ and $84^\circ$ and remains a good fit to the simple S-shaped phases at other baselines.
    Some of the remaining residuals could be caused by higher-order velocity structures beyond the scope of our kinematic model, but that could be reproduced by more complex kinematic codes.\\
    A significant advantage that our disk-wind model has over the simple Keplerian model is that it allows us to reconcile the position angle of the major axis of the continuum disk and the position angle of the "kinematic axis" measured in the AMBER and CRIRES photocenter shifts.
    Due to the position angle shift we see in our disk-wind model (see Fig. \ref{fig:model-angles} and discussion above), the $\theta$ value for our disk-wind model is within the standard deviation of the PA observed by PIONIER \citep{2017Lazareff} and AMBER LR (in contrast to the Keplerian model). Furthermore, the much better fit of the disk-wind model to the geometry of the continuum-emitting disk makes it more plausible than the Keplerian alternative.\\
    
    A disk-wind model has already been invoked by \citet{2011Weigelt} to model AMBER HR-K visibilities and phases measured on MWC297 as well as the spectral line profile.
    However, their AMBER data probed only a single position angle and a limited baseline length range (14-42\,m), but was still used to constrain simultaneously the disk orientation as well as a complex 2D velocity field.
    \citet{2011Weigelt} found that their two disk-wind models with PAs of the projected polar axis of both $65$ and $300^\circ$ are approximately able to reproduce their observations (Sect.~4.4 and Fig.~7 in their paper).
    Using their model images we computed visibilities and differential phases for our new AMBER HR-K observations with better $uv$-coverage and show them in Fig.~\ref{fig:W11-model}. 
    Adopting their best-fit PA value of the polar axis of the disk ($300^\circ$), we find that the model provides reasonable agreement to the observed visibilities with a reduced $\chi^2_{r,V} = 1.01$; although the shape of the visibility drop is systematically more narrow in wavelength than what is observed in our new data.
    However, the differential phases predicted by the model (with values up to $\pm 180^\circ$) are much stronger than observed (phases $\gtrsim \pm60^\circ$), which leads to a poor $\chi^2_{r,\phi}$ of 10.1.
    Additionally, with a best-fit minor-axis PA of $300^\circ$ ,  the \citet{2011Weigelt} model seems incompatible with measurements of the disk continuum geometry, both from PIONIER \citep[$13.7\pm1.7^\circ$,][]{2017Lazareff} and our AMBER LR measurements ($9.6\pm4.8^\circ$).
    When we orient the model to match the measured continuum geometry of MWC297 (by adopting a PA value of $10^\circ$) we find that the $\chi^2_r$ value increases to $12.8$ ($\chi^2_{r,V} = 1.33$, $\chi^2_{r,\phi} = 11.5$).
    Possibly, this significant difference between the measured disk geometry and the model orientation is linked to the effect that we explore previously in this section (and that are shown in Fig.~\ref{fig:model-angles}) where the out-of-plane velocity component distorts the perceived rotation angle.
    Therefore, this might indicate that the toroidal velocity component in the \citet{2011Weigelt} model is overestimated, causing the perceived axis of motion to differ from the model's continuum axis more than observed.
    A comparison of the different kinematic modeling codes and their various strengths will be the focus of a future study.


\section{Conclusions}

In this paper we investigated the kinematics of the Br$\gamma$-emitting gas at milliarcsecond scales with spectrally dispersed interferometry and spectro-astrometry.
Our VLTI/AMBER ($R$=$12,000$) observations represents the most extensive data set that has so far been employed for studying the gas kinematics in the inner few AU around a young stellar object, while our VLT/CRIRES ($R$=$100,000$) data allow us to measure the detailed line profile and the photocenter displacements in the line with very high SNR.

Analyzing the combined data set allows us to draw the following conclusions:
  \begin{enumerate}
  \item Our AMBER observations at low spectral dispersion ($R=35$) show that the NIR continuum is small (FWHM = $4.93$mas), occupying a region ${\sim}3.6\times$ more compact than the expected ${\sim}3$au dust sublimation radius for MWC297 (at ${\sim}17.6$mas).
  From the AMBER visibility measurements we determine that the disk has a major axis PA of $99.6^\circ \pm 4.8^\circ$ and an inclination of $32^\circ \pm 3^\circ$, consistent with the recent PIONIER observations by \citet{2017Lazareff}.
  We model the wavelength-dependent visibilities from our AMBER HR observations across the Br$\gamma$ line and find that the line-emitting region is ${\sim}2.3$ times more extended than the compact K-band continuum.
  
  \item The 2D photocenter displacement vectors derived from the AMBER and CRIRES data indicate a velocity field that is dominated by rotation.  We are able to place strong constraints on the position angle of the axis of motion and find consistent values for AMBER ($112^\circ\pm10^\circ$) and CRIRES ($114^\circ\pm3^\circ$).
  At low velocities, both data sets show a photocenter displacement towards the observer, indicative of the furthest parts of the disk being obscured by a strong milliarcsecond scale opacity gradient.
  
  \item Our model-independent channel maps resolve the Br$\gamma$ line at a spectral resolution of 12,000 and an angular resolution of 2\,mas and represent the first velocity-resolved image obtained for a young star at infrared wavelengths.
  The individual channel maps as well as the moment map indicate a rotation-dominated velocity field and show brighter emission from the northern side of the disk, consistent with the previously discussed opacity gradient.
  
  \item In order to quantify the velocity field, we fit a Keplerian-disk model and find that this model provides a moderate fit to the observed visibilities, differential phases and spectro-astrometric signal.
  However, the derived gas-disk PA in this model is inconsistent with the orientation of the continuum disk on a $3\sigma$-level, indicating the presence of a non-rotational velocity component. 
  
  \item In order to explain the significant PA difference between line photocenters and the continuum disk ($\Delta\theta=14\pm3^{\circ}$), we extend the Keplerian disk model by including a poloidal velocity component, simulating a parameterized disk-wind model.
  We find that the discrepancies between the measured line and continuum PAs can be reconciled by a disk-wind model with an out-of-plane velocity of $0.14 \times v_k$.
  Our parametric disk-wind model is able to fit the Br$\gamma$-line profile, as well as the high-resolution spectro-interferometry and spectro-astrometry data.
  
  \item Through our disk-wind modeling, we discover that the PA difference between the disk major axis and the axis of motion $\Delta\theta$ constitutes a powerful diagnostic for detecting non-Keplerian velocity contributions, which could be exploited in future observational studies.
\end{enumerate}

\begin{acknowledgements}
We thank Larisa Tambovtseva and Vladimir Grinin for helpful discussions regarding the modeling of disk winds.
We acknowledge support from a STFC Rutherford Fellowship (ST/J004030/1), Rutherford Grant (ST/K003445/1), Philip Leverhulme Prize (PLP-2013-110), and ERC Starting Grant (Grant Agreement No.\ 639889).

\end{acknowledgements}

\bibliographystyle{aa}
\bibliography{MWC297-aa}

\begin{appendix}

\section{Comparison with a previously published model}

\begin{figure*}
\centering
\includegraphics[width=18.4cm]{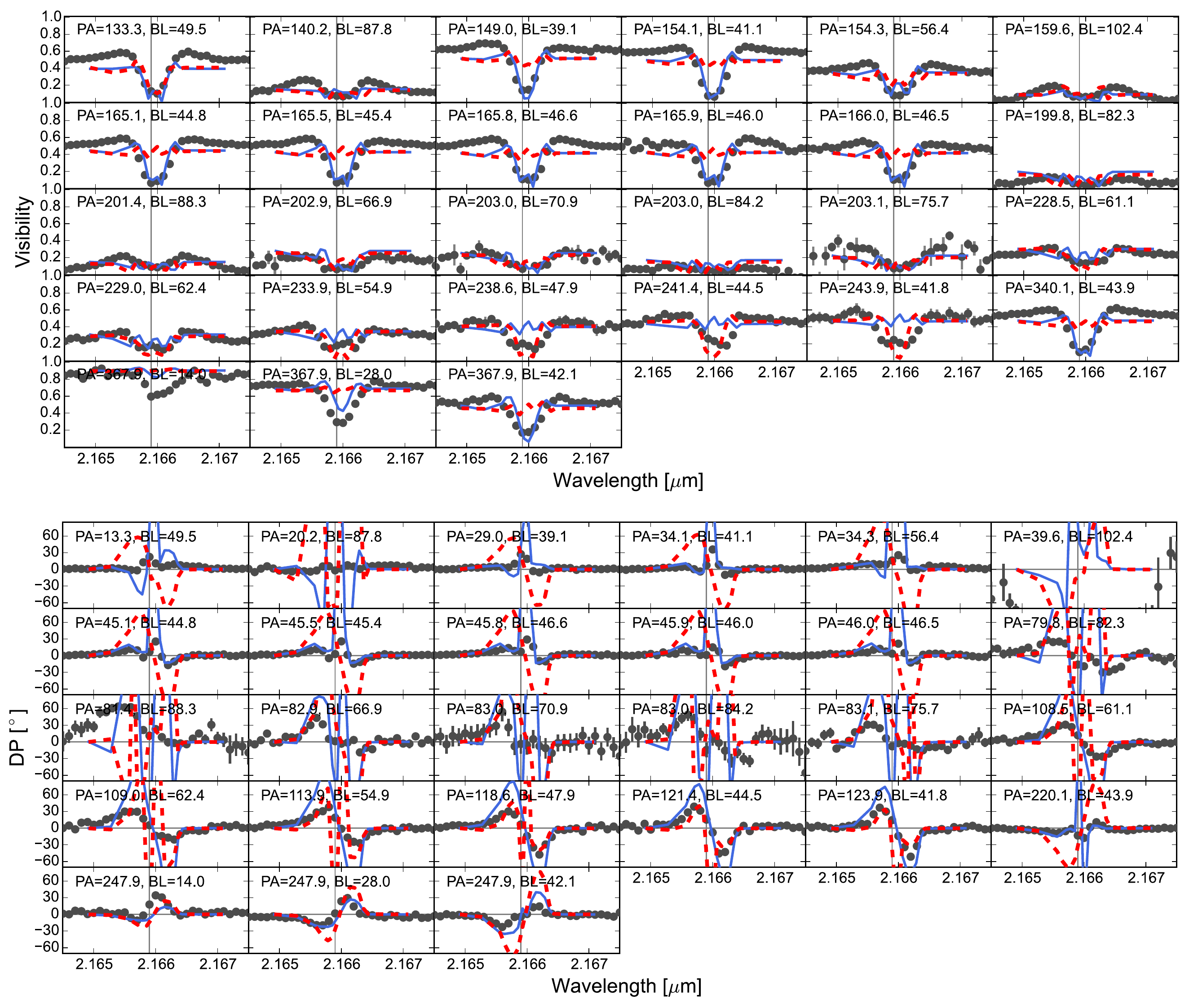}
\caption{
Visibilities (upper panel) and differential phases (lower panel) calculated from the disk wind model presented by \citet{2011Weigelt}.
The solid blue line represents the best fit model ($\theta=300^\circ$) and the dashed red line represents the model with the best match to the continuum geometry ($\theta=10^\circ$).
Data points are gray.
}
\label{fig:W11-model}
\end{figure*}

In order to compare the model from \citet{2011Weigelt} with our new interferometric data we extract the interferometric observables from their radiative transport model images.
The comparison of the visibility and phases with our data is shown in Fig.~\ref{fig:W11-model}.

\end{appendix}

\end{document}